\documentclass[12pt,a4paper]{article}
\pdfoutput=1
\usepackage{jheppub}
\usepackage{amsfonts}
\usepackage{bbm}
\usepackage{graphicx}
\usepackage[utf8]{inputenc}
\usepackage[T1]{fontenc}
\usepackage{amssymb,amsfonts,amsmath}
\usepackage{subfig} 
\usepackage{empheq}
\usepackage{upgreek} 
\usepackage{slashed}

\usepackage{tikz, pgf}
\usetikzlibrary{shapes.misc}
\usetikzlibrary{shapes}

\makeatletter

\pgfdeclareshape{genus}{
 \anchor{center}{\pgfpointorigin}
\backgroundpath{
    \begingroup
    \tikz@mode
    \iftikz@mode@fill

         \pgfpathmoveto{\pgfqpoint{-0.78cm}{-.17cm}}
     \pgfpathcurveto %
            {\pgfpoint{-0.35cm}{-.44cm}}
        {\pgfpoint{0.35cm}{-.44cm}}
        {\pgfpoint{.78cm}{-0.17cm}} 
     \pgfpathmoveto{\pgfqpoint{-0.78cm}{-0.17cm}}
     \pgfpathcurveto %
            {\pgfpoint{-0.25cm}{.25cm}}
        {\pgfpoint{.25cm}{.25cm}}
        {\pgfpoint{0.78cm}{-0.17cm}}
        \pgfusepath{fill}
        \fi

        \iftikz@mode@draw
     \pgfpathmoveto{\pgfqpoint{-1cm}{0cm}}
     \pgfpathcurveto %
            {\pgfpoint{-0.5cm}{-.5cm}}
        {\pgfpoint{0.5cm}{-.5cm}}
        {\pgfpoint{1cm}{0cm}}

     \pgfpathmoveto{\pgfqpoint{-0.75cm}{-0.15cm}}
     \pgfpathcurveto %
            {\pgfpoint{-0.25cm}{.25cm}}
        {\pgfpoint{.25cm}{.25cm}}
        {\pgfpoint{0.75cm}{-0.15cm}}
              \pgfusepath{stroke}
        \fi
        \endgroup
    }
    }

    \makeatother

\newcommand\widebar[1]{\mathop{\overline{#1}}}
\renewcommand\mod{\textrm{mod}}

\def\iimg{ { i}} 
\def\erfc{ \textrm{erfc}}
\def\wsa{\alpha}

\def\ssa{A}
\def\ssb{B}
\def\tsa{i}
\def\tsb{j}

\def\etaaps{ \eta_{_\aps} }

\def\aps{ \textrm{APS}}

\def\ln{ \textrm{ln}}

\def\spartial{\mathcal{D}}

\def\erfc{ \textrm{erfc}}

\def\scharge{Q}

\font\cmss=cmss10
\font\cmsss=cmss10 at 7pt
\font\manual=manfnt

\newcommand{\bi}{\begin{itemize}}
\newcommand{\ei}{\end{itemize}}

\newcommand{\bea}{\begin{eqnarray}}
\newcommand{\eea}{\end{eqnarray}}
\newcommand{\be}{\begin{equation}}
\newcommand{\ee}{\end{equation}}
\newcommand{\ben}{\begin{eqnarray*}}
\newcommand{\een}{\end{eqnarray*}}
\newcommand{\bem}{\begin{pmatrix}}
\newcommand{\eem}{\end{pmatrix}}
\newcommand{\bl}{\begin{align}}
\newcommand{\el}{\end{align}}
\newcommand{\beg}{\begin{gather}}
\newcommand{\eeg}{\end{gather}}

\newcommand{\erf}{ \textrm{erf}}
\renewcommand{\=}{\, = \,}
 
\newenvironment{myenumerate}{
\begin{enumerate}
   \setlength{\itemsep}{1pt}
   \setlength{\parskip}{0pt}
   \setlength{\parsep}{0pt}}{\end{enumerate}}

\newcommand{\cA}{\mathcal{A}}
\newcommand{\cB}{\mathcal{B}}

\newcommand{\cH}{\mathcal{H}}

\newcommand{\cI}{\mathcal{I}}

\newcommand{\cM}{\mathcal{M}}
\newcommand{\cN}{\mathcal{N}}

\newcommand{\cS}{\mathcal{S}}

\newcommand{\bH}{\ensuremath{\mathbb{H}}}
\newcommand{\bI}{\ensuremath{\mathbb{I}}}

\newcommand{\bQ}{\ensuremath{\mathbb{Q}}}
\newcommand{\bR}{\ensuremath{\mathbb{R}}}

\newcommand{\bZ}{\ensuremath{\mathbb{Z}}}

\newcommand{\hW}{{\widehat W}}
\newcommand{\hM}{{\widehat {\cal M}}}

\newcommand{\IH}{\mathbb{H}}

\renewcommand{\a}{\alpha}
\renewcommand{\b}{\beta}
\renewcommand{\d}{\delta}
\newcommand{\e}{\epsilon}
               
\newcommand{\g}{\gamma}

\renewcommand{\l}{\lambda}
\newcommand{\m}{\mu}

\newcommand{\s}{\sigma}                                   
\renewcommand{\t}{\tau}

\newcommand{\G}{\Gamma}

\newcommand{\inn}{{\,\in\,}}

\newcommand{\half}{\frac{1}{2}}
\newcommand{\pa}{\partial}

\newcommand{\Tr}{\mbox{Tr}}
\newcommand{\tr}{\mbox{Tr}}
\newcommand{\sgn}{\mbox{sgn}}

\renewcommand{\c}{\cdot}
\newcommand{\+}{{\,+ \,}}

\newcommand{\TrH[1]}{ {\raise -.5em
                      \hbox{$\buildrel {\textstyle  {\rm Tr } }\over
{\scriptscriptstyle \cH _ {#1}}$}~}}
\newcommand{\TrS}{ {\raise -.5em
                      \hbox{$\buildrel {\textstyle  {\rm Tr } }\over
{\scriptscriptstyle SP}$}~}}
\newcommand{\TrSb}{ {\raise -.5em
                      \hbox{$\buildrel {\textstyle  {\rm Tr } }\over
{\scriptscriptstyle SP_{b}}$}~}}
\newcommand{\TrSs}{ {\raise -.5em
                      \hbox{$\buildrel {\textstyle  {\rm Tr } }\over
{\scriptscriptstyle SP_{s}}$}~}}

\newcommand{\res[1]}{{\raise -.5em 
\hbox{$\buildrel{\textstyle{\rm Res}}\over {\scriptscriptstyle {#1}}$}}}
\newcommand{\tends[1]}{{\raise -.5em 
\hbox{$\buildrel{\longrightarrow}\over {\scriptscriptstyle {#1}}$}}}

\def\dbend{\lower3.5pt\hbox{\manual\char127}}

\def\IL{\relax{\rm I\kern-.18em L}}
\def\IH{\relax{\rm I\kern-.18em H}}
\def\rlx{\relax\leavevmode}

\def\ZZ{\rlx\leavevmode\ifmmode\mathchoice{\hbox{\cmss Z\kern-.4em Z}}
 {\hbox{\cmss Z\kern-.4em Z}}{\lower.9pt\hbox{\cmsss Z\kern-.36em Z}}
 {\lower1.2pt\hbox{\cmsss Z\kern-.36em Z}}\else{\cmss Z\kern-.4em
 Z}\fi}

\title{ APS  $\eta$-invariant, path integrals, \\
and mock modularity}

\author[1, 3, 4]{Atish Dabholkar,}
\author[1, 2]{Diksha Jain,}
\author[1]{Arnab Rudra\vspace{10pt}}

\affiliation[1]{International Centre for Theoretical Physics\\
Strada Costiera 11, Trieste 34151 Italy
\vspace{4pt}}

\affiliation[2]{SISSA, Via Bonomea 265, Trieste 34136 Italy
\vspace{4pt}}

\affiliation[3]{  Sorbonne Université, CNRS, \\
Laboratoire de Physique Théorique et Hautes Énergies, \\LPTHE, F-75005 Paris, France
\vspace{4pt}}

\affiliation[4]{CNRS, UMR 7589, LPTHE,  Paris, F-75005 France
\vspace{4pt}}

\abstract{We  show  that the Atiyah-Patodi-Singer $\eta$-invariant can be related to the temperature dependent Witten index of a noncompact theory and give a new proof of the APS theorem using scattering theory. We relate the $\eta$-invariant   to a Callias index and compute it using localization of a supersymmetric path integral. We show that the $\eta$-invariant for the elliptic genus of a finite cigar is related to  quantum modular forms obtained from  the completion of a mock Jacobi form which we compute from the noncompact path integral.\\
\leftline{}
\centerline{(\textit{Dedicated to the memory of Michael Atiyah})}
}

\begin{document}
\maketitle

\hspace{10 cm}

\centerline{\it``Any good theorem should have several proofs, the more the better.''}
\vspace{5mm}
\centerline{\hspace{6cm}\sl Michael Atiyah}	

%
%

\section{Introduction}

In this note, we establish a connection between three distinct sets of ideas— 
the Atiyah-Patodi-Singer $\eta$-invariant \cite{Atiyah:1975jf},  mock Jacobi forms \cite{Dabholkar:2012nd} and associated quantum modular forms \cite{zagier2010quantum}, and supersymmetric path integrals. The link between these three is provided by the \textit{temperature-dependent} Witten index of a noncompact theory.
For a supersymmetric quantum field theory in a $d$-dimensional spacetime, the Witten index \cite{Witten:1982im} is defined by
\be\label{witten}
W(\b) := \TrH[] \left[  (-1)^{F} e^{-\beta H} \right]
\ee 
where $\beta$ is the inverse temperature, $H$ is the Hamiltonian, $F$ is the fermion number and $\cH$ is the Hilbert space of the theory. 
As usual, this trace can be related to a  supersymmetric  Euclidean path integral over a $d$-dimensional Euclidean base space $\Sigma$ with periodic boundary conditions for all fields in the Euclidean time direction.

If the quantum field theory is compact in the sense that the spectrum of the Hamiltonian is discrete, then the Witten index is independent of the inverse temperature $\b$. This follows from the observation that the states with nonzero energy come in Bose-Fermi pairs and do not contribute to the Witten index \cite{Witten:1982im}. Only the zero energy states graded by $(-1)^F$ contribute and consequently, the Witten index is a  topological invariant. This is the case, for example, for a supersymmetric sigma model with a compact target space. By an appropriate choice of the sigma model, the Witten index in the zero temperature  ($\beta \rightarrow \infty$) limit can be related to some of the classic topological invariants such as the Euler character or the Dirac index of the target manifold.  Using temperature independence of the index, one can evaluate it in the much simpler high temperature   ($ \beta \rightarrow 0$) limit using the heat kernel expansion to prove the Atiyah-Singer index theorem \cite{Atiyah:1963zz}. Evaluating the path integral corresponding to the Witten index in this high-temperature semiclassical limit gives another derivation of the index theorem \cite{AlvarezGaume:1983rm,Friedan:1983xr}.

If the field space is noncompact and the spectrum is continuous, then the above argument can fail because now instead of a discrete indexed sum, one has an integral over a continuum of scattering states. To define the noncompact Witten index properly,  one needs a framework to incorporate the non-normalizable scattering states into the trace. We address this issue in \S\ref{Noncompact} and give a suitable definition using the formalism of Gel'fand triplet.  In general, the bosonic density of states in this continuum may not precisely cancel the fermionic density of states, and the noncompact Witten index can be temperature dependent. 
We relate this temperature dependence to the Atiyah-Patodi-Singer $\eta$-invariant and mock modularity and compute it using deformation invariance and localization of the supersymmetric path integral. The temperature dependent piece is no longer topological but is nevertheless  `semi-topological' in that it is independent of any deformations that do not change the asymptotic.  Note that in a general non-compact situation, temperature independence and deformation invariance are logically distinct as will become clear later. 

The relation to the APS $\eta$-invariant can be seen as follows. Consider a compact manifold $\cM$ with a boundary $\cN$.  One can define Atiyah-Patodi-Singer boundary conditions \cite{Atiyah:1975jf}   for the Dirac operator  assuming   that $\cM$ has a product form  near the boundary $\cM \sim    \cN \times \bI \, $
where $\bI$ is a finite interval. It is useful to introduce a noncompact manifold $\widehat \cM$ obtained by a trivial extension of the manifold  $\cM$  by gluing it to a half-cylinder $\cN \times \bR^{+} $.  One can define the Witten index $\hW(\b)$ where the hat is added as a reminder that it corresponds to the noncompact theory with $\widehat \cM$ as the target space. By construction, the solutions to the Dirac operator on $\cM$  with  APS boundary conditions are in one-one correspondence with the solutions to the Dirac operator on the $\widehat\cM$ with the square-integrable norm.  However,  the spectrum of the Dirac Hamiltonian on $\widehat \cM$  also contains delta-function normalizable scattering states with continuous energies, in addition to  the square-integrable bound states with discrete energies. We assume that the continuum is separated from the ground states by a gap. 
In the $\b \rightarrow \infty$ limit, only the ground states corresponding to the square-integrable solutions of the Dirac operator contribute and hence the APS index for the original compact manifold $\cM$  equals $\hW(\infty)$.   There is a bulk contribution to the APS index coming from the integral of a local index density which can be computed in the  $\beta \rightarrow 0$ limit as in the compact case.  However, now there is a left-over piece coming from the contribution of the continuum. Using  scattering theory on $\widehat \cM$ we relate the continuum contribution to the $\eta$-invariant of $\cN$ by the relation 
\be\label{etawitten}
\eta  = 2 \left(\hW(0) - \hW(\infty) \right) \, .
\ee
This yields a new proof of the APS theorem. 
The $\eta$-invariant is nonzero precisely because the noncompact Witten index is temperature dependent.  

The relation to mock modularity arises similarly as a consequence of noncompactness of the field manifold for a superconformal field theory on a base space $\Sigma$ which is a 2-torus with a complex structure parameter $\tau$. The elliptic genus of an SCFT is a generalization of the  Witten index that counts the right-moving ground states with arbitrary left-moving excitations. It  is \textit{a priori} a function of $\tau$ and $\bar\t$. For a compact SCFT, by an argument similar to the above, it is independent of the `right-moving temperature' and hence of $\bar \t$, and is a (weakly)  holomorphic Jacobi form. Once again, for a noncompact SCFT with $\widehat \cM$ as the target space, this argument fails.  There is a `holomorphic anomaly' because the right-moving bosonic density of states does not precisely cancel the right-moving fermionic density of states. 
In this case, the elliptic genus is no longer a Jacobi form but is rather a completion of a mock Jacobi form—a new mathematical object introduced in \cite{Dabholkar:2012nd}. The holomorphic anomaly is once again governed by the temperature dependence of the `noncompact right-moving Witten index'. 

An advantage of mapping the APS index to the Witten index on the noncompact manifold $\widehat \cM$  is that it becomes easier to obtain its path integral representation. 
  Defining a path integral measure in a target space with a boundary is in general rather complicated. Even for a very simple system like a particle in a box, the path integral formulation was achieved relatively recently {\cite{Clark:1980xt, Goodman:1981abc, Farhi:1989jz, Inomata:1980th, Carreau:1990wh, Carreau:1991yx}. For a path integral on a  manifold $\hM$ without a boundary, even if it is noncompact,  one can use the canonical measure.   The path integral facilitates computations using supersymmetric localization. We derive the APS result by relating it to a Callias-Bott-Seeley \cite{Callias:1977kg, Bott:1978bw} index theorem as we explain in \S\ref{Callias}.  
 A path integral representation also makes the modular invariance manifest making it easier to see the connection with mock modularity. 

Apart from its intrinsic importance  in differential topology, the  $\eta$-invariant has a number of interesting physics applications, for example,  in the analysis of  global gravitational anomalies \cite{Witten:1985xe}, in fermion fractionization \cite{Niemi:1984vz,Lott:1984tn} , in relation to spectral flow in quantum chromodynamics \cite{Callan:1977gz,Kiskis:1978tb}, and more recently  in the description of symmetry-protected phases of topological insulators (see \cite{Witten:2015aba} for a recent review). Similarly, apart from  their intrinsic interest in   number theory \cite{Zwegers:2008zna,MR2605321}, mock modular forms and their cousins have  come to play an important role in the physics of quantum black holes and quantum holography \cite{Dabholkar:2012nd}, in umbral moonshine \cite{Eguchi:2010ej,Cheng:2012tq}, in the context of WRT invariants \cite{Witten:1988hf, Reshetikhin:1991tc, lawrence1999modular, Cheng:2018vpl}
, and more generally in the context of elliptic genera of noncompact SCFTs \cite{Eguchi:2010cb, Troost:2010ud, Sugawara:2011vg, Ashok:2013pya,Murthy:2013mya, Giveon:2015raa}. We expect our results will have useful implications in these diverse contexts. 

The paper is organized as follows. 
In \S\ref{Compact} we review the supersymmetric quantum mechanics for index theory on a compact manifold without a boundary 
and in \S\ref{APS} we describe the Atiyah-Patodi-Singer construction for a compact manifold with a boundary.
 In \S\ref{Elliptic} we present the definition of the elliptic genus of an SCFT and its relation to Jacobi forms. 
In  \S\ref{Noncompact} we define the noncompact Witten index using the Gel'fand triplet. Then we use this formalism to present a proof of the APS theorem in \S\ref{Scattering}. 
In \S\ref{Path}  we discuss the path integral representation 
 of the $\eta$-invariant. In \S\ref{Callias} we apply localization to reduce the path integral for the $\eta$-invariant to an ordinary super-integral evaluated in \S\ref{Worldpoint} and relate it to the Callias index. We use these results to compute the $\eta$-invariant of the finite cigar in \S\ref{Cigar} and to compute the elliptic genus for the infinite cigar in \S\ref{CigarEll}.  We review the definitions of mock Jacobi forms in \S\ref{MockJ} and discuss the connection with the $\eta$-invariant and quantum modular forms in  \S\ref{Quantum}.

The examples considered in this paper are simple but sufficiently nontrivial and illustrative.  
Our results indicate that these interesting connections are a rather general consequence of noncompactness.  Supersymmetric methods have been used successfully to obtain a path integral derivation of the Atiyah-Singer index theorem for a compact target manifold, but to our knowledge, no such derivation exists for a manifold with a boundary\footnote{Indeed, this was posed by Atiyah a decade ago as a problem for the future \cite{AtiyahBott60}.}.  Using our formulation in terms of a noncompact Witten index, it should be possible to obtain a more complete path integral derivation of the APS index theorem,  for example, even for the manifolds that do not have product form near the boundary \cite{GILKEY1975334}. It would also be interesting to generalize the construction to elliptic genera of generic noncompact superconformal field theories. We will return to these problems in the future. 

\section{Supersymmetry and index theorems \label{Susy}}

We will be interested in  supersymmetric path integrals for $d$-dimensional quantum field theories with $d=2, 1, 0$. The Euclidean base space $\Sigma$ in the three cases is  a 2-torus $T^{2}$, a circle $S^{1}$, and a   point which we will refer to as the worldsheet, worldline, and worldpoint
 respectively. All our examples are obtained by  reductions of  Euclidean Wick-rotated version of a  $1+1$ dimensional worldsheet  with $(1, 1)$ supersymmetry which we describe below.  It is convenient to use the superspace $s\Sigma$ with real superspace coordinates $\{ \s^{\a} , \theta_{A}\}$.  See \S\ref{Spinor} for the conventions.
We  write $\sigma^{1}= \s$ and $\sigma^{0}=t$ or $ \sigma^{0}= \t$  in the Lorentzian or Euclidean version respectively. 

Let  $\{X^i(\s, \theta) \}$  be  real super-fields with  expansion
\begin{eqnarray}\label{sfield}
X^i (\sigma) =x^i(\s) +\bar \theta \psi^i  (\s)+\frac{1}{2}\bar \theta \theta F^i (\s)
\label{twodimsusy31}
\end{eqnarray}
where  $\{x^{i}\}$ are the coordinates of  $2n$ dimensional real field manifold $\cM$, 
 $\psi^i_{A}$ are real Grassmann  fields and $F^i$ are  auxiliary fields. 
The components of the superfield can   be thought of as the coordinates of a supermanifold $s\cM$. 
The Lorentzian action  is
\begin{eqnarray}\label{sheet}
I(X)&=& - \frac{1}{2 \pi \alpha'}
\int_{s\Sigma}	d^2 \s\, d^2\theta \,    \left[ \frac{1}{2} g_{ij}(X) \bar \spartial  X^i \spartial X^j + 2 \, h(X) \right]
\end{eqnarray}
where  $g_{ij}(x)$ is the metric on $\cM$, $h(X)$ is the superpotential, and $\spartial$ is the superspace covariant derivative on the (base) superspace. We have introduced $\alpha'$ for easy comparison with other normalizations in the literature.
The action is invariant under diffeomorphisms in the target space $\cM$. It is also invariant under translations of $t$ generated by the Hamiltonian $H$ and translations of $\sigma$ generated by $P$ as well as  under the $\bZ_{2}$ action of $(-1)^{F}$:
\be
\psi^i{ \rightarrow} - \psi^i \, ,  \qquad x^{i} \rightarrow x^{i} \, .
\ee
Moreover, it is invariant under the $(1, 1)$  supersymmetry generated by
a real constant spinor $\e_{A}$  under which  the superfield transforms as $\delta X = (\bar \e Q )X $ and its components transform as
\begin{eqnarray}
\delta 	x^i &=& \bar \epsilon \psi^i 
\nonumber\\
\delta 	\psi^i &=&(-\iimg \gamma^\a  \partial_\a x^i+F^i)\epsilon
\\
\delta 	F^i &=&-\iimg \bar  \epsilon\gamma^\a   \partial_\a \psi^i \, . \nonumber 
\end{eqnarray}
With $\alpha' = 1$, the action \eqref{sheet} in superfield components is given by
\begin{eqnarray}\label{action2}
I &=& -\frac{1}{2 \pi}\int_{\Sigma}  d^{2}\sigma   \Big[ \frac{1}{2}g_{ij}\left(
\partial_\alpha x^i\partial^\alpha x^j
- \iimg \bar \psi^i 
\slashed{\nabla}\psi^j 
-F^iF^j
\right)	
\nonumber\\
&&
\, +\, \frac{1}{4}\partial_k\partial_l g_{ij}(x) (\bar \psi^k  \psi^l )(\bar \psi^i \psi^j)
\, -\,  \frac{1}{4}\partial_k g_{ij}\bar \psi^i \psi^j F^k+\frac{1}{4}\partial_k g_{ij}(F^i\bar \psi^j+F^j\bar \psi^i )\psi^k 
\\
&&
\, +  \frac{\partial h}{\partial x^i} F^i
- \frac{1}{2}
\frac{\partial^2 h}{\partial x^i\partial x^j}( \bar\psi^i \psi^j) \Big] \nonumber
\label{susysigma5} 
\end{eqnarray}
where the covariant derivative 
\begin{eqnarray}
\nabla_\a \psi^i=\partial_\a \psi^i+{\Gamma^i}_{jk}\, 	\partial_\a x^j\psi^k 
\label{susysigma7} 
\end{eqnarray}
is defined using the Christoffel symbols $\G^{i}_{jk}(x)$ in the target space. 
When the superpotential is zero, eliminating the auxiliary fields yield the familiar quartic fermionic term involving the Riemann curvature tensor \cite{AlvarezGaume:1983ab,Gates:1983py}. 
It is convenient to introduce an orthonormal basis of forms, 
$ e^{a} = e^{a}_{\, i} dx^{i}$,  using the vielbein  $e^{a}_{\, i}$ and the inverse vielbein $e^{i}_{\,a}$ with  $1 \leq a, b \leq 2n$ as the tangent space indices. The metric can then be expressed as $g_{ij}=  e^{ a}_{\, i}e^{ b}_{\, j}\d_{ab}$, and  one can define the spin connection ${\omega^{a}}_{kb}$ associated with the Christoffel symbols.  

The target space $\cM$  may be compact with or without boundary, or noncompact. As explained in the introduction, each of these cases have to be treated differently with an appropriate definition of the Witten index.

\subsection{Compact Witten index \label{Compact}}

To set the stage, we first consider a  superparticle on  compact  $\cM$ used in the famous  \cite{Witten:1982im, AlvarezGaume:1983rm,Friedan:1983xr} path-integral derivation of the Atiyah-Singer index theorem.  The Lorentzian action
is given by 
\begin{eqnarray}\label{point}
I = \frac{1}{2} \int \, dt  \left[  g_{ij }(x)\frac{dx^i }{d t }\frac{dx^j }{d t }
+  i \, \delta_{ab}\psi^a\left( 
\frac{d\psi^b }{d t}
+ \omega_{akb}
\frac{dx^k }{d t }
\psi^b 
\right)\right ]\, ,
\end{eqnarray}
which can be obtained   as a specialization of \eqref{sheet} by setting
\be
F^{i}=0 \, , \qquad  h =0 \,, \qquad \psi_{-}^{a} =0 \, , \qquad \frac{\partial}{\partial \s} =0 \, , \qquad \psi_{+}^{a} = \psi^{a} \, .
\ee
We have defined  $\psi^{a} =  e^{ a}_{\, i}  \psi^{i}$ using the vielbein.  The conjugate variables are
\be
\pi_{a} : = \frac{ \partial L}{\partial \dot \psi^{a}}=  \frac{i}{2} \psi_{a} \, \qquad p_{i} : = \frac{ \partial L}{\partial \dot x^{i}}= \dot x_{i} + \frac{i}{2} \psi^{a}\omega_{iab} \psi^b 
\ee
where  the dot refers to $t$-derivative. The nonvanishing canonical commutation relations are\footnote{The naive anticommutator obtained from the Poisson bracket  of $\pi$ and $\psi$ would imply 
$\{\psi^{a}, \psi^{b}\} = 2\delta^{ab}$ apparently in conflict with the anticommutator of  $\psi$ with itself,  $\{\psi^{a}, \psi^{b}\} =0$.  However, since $\pi$ is proportional to $\psi$, we have a constraint on the phase space and must use Dirac brackets instead of Poisson brackets to obtain the correct quantization   $\{\psi^{a}, \psi^{b}\} = \delta^{ab}$, roughly as an average of the naive commutators. }
\be
    \{\psi^a, \psi^b\} =  \delta ^{ab} \, , \qquad  [x^{i}, p_{j}] = i \delta^{i}_{j} \, .
\ee
The Hilbert space $\cH$  furnishes a Dirac representation of $2n$-dimensional  $\gamma$-matrices with $\sqrt{2}\psi^j =-\iimg  \gamma^j$. 
The chirality matrix $\bar \g$ for the Dirac representation 
\be
\bar \gamma = i^{n} \gamma^{1}\gamma^{2}\ldots \gamma^{2n} \, , \quad (\bar \gamma)^{2} =1\, 
\ee
can be identified with $(-1)^{F}$. For a review see \cite{Nakahara:2003nw}  which uses  slightly different conventions.

The worldline supersymmetry is  now parametrized by a single Grassman parameter $\e = -\e_{-}$
\begin{eqnarray}
	\delta x^i = i \epsilon \psi^i \, , \qquad \delta \psi^i = - \epsilon\dot{x}^i \, .
\end{eqnarray}
The corresponding Noether supercharge is
\begin{eqnarray}
    \epsilon Q &=& -\epsilon \sqrt{2} \psi^i \dot{x}_i \, ,
\end{eqnarray}
Upon quantization, we get
\begin{eqnarray}
    Q =    \gamma^i D_{i}  = \slashed{D} \, \qquad \textrm{with} \qquad D_{i} = \partial_{i}+ \frac{1}{4} \omega_{iab} \g^{a}\g^b \, .
    \label{sc1} 
\end{eqnarray}
 which is the Dirac operator on manifold $\cM$. The canonical commutations imply the commutation relations
\begin{eqnarray}\label{anticomm}
    \{ Q, Q\} =  2 H \, , \qquad  [ H, Q] =0 \, , \qquad \{ Q, (-1)^{F}\} =0 \, ,  \qquad \{ \bar\g, \g^{a}\} =0
\end{eqnarray}
where $H$ is the worldline Hamiltonian. 
In the basis in which $\bar \g$ is diagonal, the Dirac spinor $\Psi(x)$ on field space  can be written as 
\be
\Psi = \left(
\begin{matrix}
   \Psi_{-} \\ 
\Psi_{+}
\end{matrix}
\right) \, ,
\ee
where the $\pm$ denote the  chiralities (not to be confused with the  chirality of the Grassman field $\psi_{A}(\s)$ on the worldsheet). The Dirac operator  and the Hamiltonian take the form
\begin{eqnarray}
	 \slashed{D}=	
\left(
\begin{matrix}
   0 & L \\ 
 L^\dagger  & 0 
\end{matrix}
\right) \, , \qquad \qquad  H=	
\left(
\begin{matrix}
   L  L^\dagger  & 0 \\ 
 0  &  L^\dagger L 
\end{matrix}
\right) : = \left(
\begin{matrix}
   H_{-}  & 0 \\ 
 0  &  H_{+}\end{matrix}
\right) \,.
\label{apsindexth01}
\end{eqnarray}
Using \eqref{sc1} and \eqref{anticomm}, the Hamiltonian above can be identified with the Hamiltonian of supersymmetric quantum mechanics. The index $\cI$ of the Dirac operator on the manifold $\cM$ is then defined as:
\begin{eqnarray}
	\cI = \textrm{dim Ker} L - \textrm{dim Ker} L^\dagger 
	=  n_{+} -  n_-
\label{apsindexth02}
\end{eqnarray}
where $n_+$ is the number of zero modes of $L$ with  positive chirality and $n_-$ is the number of zero modes of $L^{\dagger}$ with negative chirality,  equivalently  the number of zero modes of the Hamiltonians $H_{+}$ and $H_{-}$ respectively i.e.
\begin{eqnarray}
\cI = \textrm{dim Ker} L^\dagger L - \textrm{dim Ker} L L^\dagger = \lim_{\beta \rightarrow \infty} \Tr (\exp^{-\beta H_+} - \exp^{-\beta H_-})
\label{eq11}
\end{eqnarray} 
Positive and negative chirality spinors on the field space correspond to positive and negative eigenstates of the operator $(-1)^{F}$ in the Hilbert space $\cH$ and can  thus be interpreted as bosonic and fermionic states  in the worldline Hilbert space $\cH$ and the index $\cI$ is naturally identified with the Witten index of the supersymmetric quantum mechanics. 
\be
\cI =  \lim_{\beta \rightarrow \infty}\TrH[] (-1)^F \exp^{-\b H} =  \lim_{\beta \rightarrow \infty} W(\b)
\ee
For a compact manifold the eigenvalues of $H$ are discrete. It then follows from \eqref{anticomm} that if $|E, +\rangle$ is a bosonic eigenstate with energy eigenvalue $E >0$, 
then $Q |E, +\rangle := |E, -\rangle$ is a fermionic eigenstate with the same energy eigenvalue. Hence,  eigenstates with non-zero eigenvalues of $H$ come in Bose-Fermi pairs and cancel out of the trace. The Witten index in this case receives contribution only from the ground states and is a  topological invariant 
\cite{Witten:1982im, Atiyah:1963zz}. 

The Witten index  
 \eqref{witten} of  this worldline theory  has a path integral representation 
\begin{eqnarray}\label{witten2}
 W(\b) &=& \int dx \,  \langle x|   (-1)^F e^{-\beta H} | x\rangle  
\qquad\qquad
 \nonumber\\
   &=& \int [dX] \, \exp\left(-S[X, \beta] \right) \, ,
\end{eqnarray}
where the path integral is over superfield configurations that are periodic in Euclidean time with period $\b$,  so the Euclidean base space $\Sigma$ is a  circle of radius $\beta$. 
The Euclidean time $\t$ is related to the Lorentzian time $t$ as usual by Wick rotation $t = - i \tau$ and 
the Euclidean action is:
\begin{eqnarray}
S[X, \b] = \frac{1}{2} \int_{0}^{\b} \, d\t  \left[  g_{ij }(x)\frac{dx^i }{d \t }\frac{dx^j }{d \t }
+  \psi^a\left( \delta_{ab}
\frac{d\psi^b }{d \t}
+ \omega_{kab}
\frac{dx^k }{d \t }
\psi^b 
\right)\right ]\, . 
\end{eqnarray}
The measure $[dX]$ is induced from the supermeasure\footnote{It is well-known that the supermeasure is  flat even if the  manifold $\cM$ is curved  because the factor of  $\sqrt{g}$ in the bosonic measure $dx := d^{2n}x \, \sqrt{g}$ cancels against a similar factor in the fermionic measure  
$d\psi := d^{2n}\psi \,  \frac{1}{\sqrt{g}}$. }
 on the supermanifold $s\cM$ introduced after \eqref{sfield}. 
One can now use the deformation invariance of the supersymmetric path integral to obtain
\be
\cI = \int_{\cM} \alpha(x)
\ee
where $\alpha(x)$ is the topological index density given by the Dirac genus.  

\subsection{Atiyah-Patodi-Singer  $\eta$-invariant \label{APS}}

Consider a  compact manifold $\mathcal{M}$ with a single boundary $\partial \mathcal{M } = \cN$ where $\cN$ is a compact, connected, oriented manifold with no boundary. 
\begin{figure}[htbp] 
\begin{center}
\begin{tikzpicture}[line width=1.5 pt, scale=.3]
\draw [blue] (-10,-1) .. 
controls (-3,-4) and (3,-4) .. 
(10,-1);
\draw [blue] (-10,-9) .. 
controls (-3,-6) and (3,-6) .. 
(10,-9);
\draw [blue] (-10,-1) .. 
controls (-20,2) and (-20,-12) .. 
(-10,-9);
\node [genus, blue, draw, ultra thick, scale=1.5] at (-11, -5) {};
\begin{scope}[shift={(5,-5)}, scale=5]	 
\filldraw[white, ultra thick] (1,1) -- (1,-1) -- (-1,-1) -- (-1,1)-- (1,1)  ;
\end{scope}
\begin{scope}[shift={(-.1,-5)}, scale=1.75]	   
\draw[red,  ultra thick] (0,1)-- (2,1)  ;
\draw[red, ultra thick] (0,-1)-- (2,-1)  ;
\end{scope}
\begin{scope}[shift={(3.2,-5)}, scale=1.75]	   
\draw [red] (0,0) ellipse (15 pt and 27pt); 
\end{scope}
\node[] at (-21,-5) {$\mathcal{M} $}; 
\end{tikzpicture}
\end{center}

\caption{Manifold $\cM$ with a  collar $\cN \times \bI$ shown in red.}\label{fig1}
\end{figure}
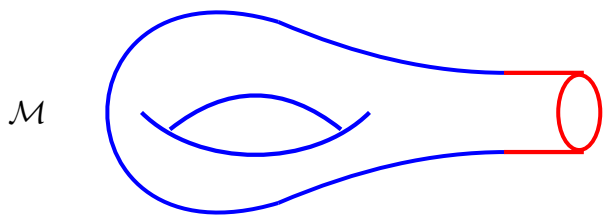
Usual local boundary conditions like Dirichlet or Neumann do define a self-adjoint Dirac operator. However, because of the reflection at the boundary, such local boundary conditions mix the positive and negative chirality and do not allow one to define the index.  To preserve chirality, it is necessary to impose the nonlocal Atiyah-Patodi-Singer boundary conditions  \cite{Atiyah:1975jf}, assuming  $\mathcal{M}$ has a  product form  in the `collar' region $\cN \times \bI$ near the boundary (Figure \textbf{\ref{fig1}}).  In local coordinates $\{y^m \,; m = 1, 2, \ldots, 2n-1\}$ on $\cN$ and $ u \leq 0 $ on  the interval $\bI$ with the boundary located at $u=0$,  the  metric takes the form
 \begin{eqnarray} \label{productmetric}
    ds^2 &=& du^2 + g_{m n}|_{\cN} \, dy^m dy^n   \, , 
\end{eqnarray}
 The Dirac operator near the boundary becomes
\begin{eqnarray}\label{apsdirac}
   \slashed{D}  =   \gamma^u \partial_u+ \gamma^m D_{m} \, .
\end{eqnarray}
It can be expressed as
\be
\label{apsdecomposition1}
 \g^{u}(\partial_u\ + \bar\gamma \cB )
\ee
where $\cB =   \widehat \gamma^m D_m$ is the boundary Dirac operator  with
$\widehat{\gamma}^m$  defined by
\be
\gamma^m =( \gamma^u \bar \gamma)	\widehat \gamma^m
\label{apsindexth41}
\ee
which satisfy the same Clifford algebra as the original $\g$ matrices:
\be
\{\gamma^m,\gamma^n\}	= -2\, g^{m n} \, ,
\qquad \qquad
\{\widehat \gamma^m,\widehat \gamma^n\}	= -2\, g^{m n} \, .
\label{apsindexth42}
\ee
The eigenvalue equation for the Dirac operator near the boundary takes the form
\begin{eqnarray}
\left(
\begin{matrix}
   0 & L  \\ 
 L^\dagger  & 0 
\end{matrix}
\right)
\left(
\begin{matrix}
   \Psi_{-} \\ 
\Psi_{+}
\end{matrix}
\right)
=
\sqrt{E}
\left(
\begin{matrix}
   \Psi_{-} \\ 
 \Psi_{+}   
\end{matrix}
\right)	
\label{apsindexth63}
\end{eqnarray}
The  eigenfunctions  can be written as  
\begin{eqnarray}
\Psi_{-} (u, y)=\sum_\lambda \Psi_{-}^{\lambda}(u)\,  e_\lambda(y)	
\label{apsindexth71a}
\\
\Psi_{+} (u, y)=\sum_\lambda \Psi_{+}^{\lambda}(u)\,  e_\lambda(y)
\label{apsindexth71b}
\end{eqnarray}
where 
$\{e_\lambda(y)\}$ are the complete set of eigenmodes of $\cB$. For each mode we obtain
\begin{eqnarray}
\left(
\frac{d}{du}+\lambda \right)	\Psi_{+}^{\lambda}(u) = \sqrt{E}  \Psi_{-}^{\lambda}(u)
\nonumber\label{apsindexth72a}
\\
\left(-
\frac{d}{du}+\lambda \right)	\Psi_{-}^{\lambda}(u) = \sqrt{E} \Psi_{+}^{\lambda}(u) \, .
\label{dirac1}
\end{eqnarray}
\begin{figure}[htbp] 
\begin{center}
\begin{tikzpicture}[line width=1.5 pt, scale=.3]
\draw [blue] (-10,-1) .. 
controls (-3,-4) and (3,-4) .. 
(10,-1);
\draw [blue] (-10,-9) .. 
controls (-3,-6) and (3,-6) .. 
(10,-9);
\draw [blue] (-10,-1) .. 
controls (-20,2) and (-20,-12) .. 
(-10,-9);
\node [genus, blue, draw, ultra thick, scale=1.5] at (-11, -5) {};
\begin{scope}[shift={(5,-5)}, scale=5]	 
\filldraw[white, ultra thick] (1,1) -- (1,-1) -- (-1,-1) -- (-1,1)-- (1,1)  ;
\end{scope}
\begin{scope}[shift={(-.1,-5)}, scale=1.75]	   
\draw[blue,  ultra thick] (0,1)-- (3,1)  ;
\draw[blue, ultra thick] (0,-1)-- (3,-1)  ;
\end{scope}
\begin{scope}[shift={(2.9,-5)}, scale=1.75]	   
\draw[red,  ultra thick] (0,1)-- (5,1)  ;
\draw[red, ultra thick] (0,-1)-- (5,-1)  ;
\draw [red] (0,0) ellipse (15 pt and 27pt); 
\end{scope}
\node[] at (-21,-5) {$\mathcal{\widehat M} $}; 
\end{tikzpicture}
\end{center}
\caption{The noncompact  $\hM$ is a  trivial extension of $\cM$  obtained by  gluing  $\cN \times \bR^{+}$. }\label{fig2}
\end{figure}
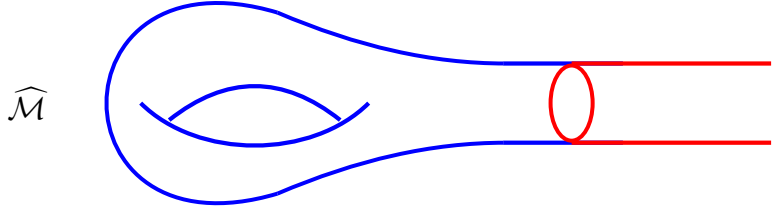
To motivate the APS boundary conditions consider a noncompact `trivial' extension $\hM$ obtained by gluing a semi-infinite cylinder $\cN \times \bR^{+}$  where $\mathbb{R}^+$ is the half line $u \geq 0$
(Figure \textbf{\ref{fig2}}). 
Near the boundary,  the zero energy solutions on $\cM$ have the form
\be
\Psi_{\pm}^{\l}(u)=\exp\, (\mp \lambda\,  u) \Psi_{\pm}^{\l}(0)	
\label{apsindexth44}
\ee
One can ask which of these solutions can be  extended to square-integrable or $L_{2}$-normalizable solutions on  the noncompact manifold $\hM$. Since $u$ is positive on the semi-infinite cylinder, the solutions are normalizable if the argument of the exponent is negative. 
This is consistent with the  APS boundary condition \cite{Atiyah:1976jg}. One sets   the  exponentially growing mode  to  zero 
which amounts to  Dirichlet boundary condition for half the modes:
\begin{eqnarray}
\Psi_{+}^{\lambda}(0) &=0 \qquad &\forall \quad \lambda < 0	
\nonumber\label{apsindexth94a}\\
\Psi_{-}^{\l}(0) &=0  \qquad &\forall \quad \lambda >0	 \, .
\label{apsindexth94b}
\end{eqnarray}
For the remaining half,  one uses 
Robin boundary conditions 
\bea
 \frac{d\Psi_{+}^{\lambda}}{du} (0) + \l
\Psi_{+}^{\lambda}(0) &=0 \qquad &\forall \quad \lambda >  0	
\nonumber\label{apsindexth94a}\\
- \frac{d\Psi_{-}^{\lambda}}{du} (0) + \l
\Psi_{-}^{\lambda}(0) &=0  \qquad &\forall \quad \lambda  < 0	 \, .
\label{apsindexth94c}
\eea
which are consistent with supersymmetry as one sees from \eqref{dirac1}. By construction, imposing the APS boundary condition on $\cM$ is equivalent to requiring $L_2$-normalizability  for the solutions of Dirac equation on the noncompact extension $\hM$.

The APS index theorem states that  the index of Dirac operator with  APS boundary conditions on the  compact Riemannian manifold $\cM$ with boundary $\cN$
  is given by 
\be\label{aps}
\cI = \int_{\mathcal{M}} \alpha (x)-\frac{1}{2}  \eta 
\ee
where  $\int_{\mathcal{M}} \alpha(x)$ is Atiyah-Singer term present also in the compact case  and $\eta$ is the Atiyah-Patodi-Singer $\eta$-invariant. It is a measure of the spectral asymmetry which is equal to the regularized  difference in the number of modes with positive and negative eigenvalues of the boundary operator $\cB$ on $\cN$. Let  $\{ \lambda \}$ be the set of eigenvalues of  $\mathcal{B}$, then 
\be
\eta = \sum_{\lambda} \sgn (\lambda) 
\qquad.
\ee
There is an ambiguity in defining the sign for $\l=0$ but one can consistently define (see the discussion before Figure \textbf{\ref{fig5}})
\bea\label{etadef}
\sgn (\lambda) &= 1  \, \qquad &\textrm{for} \qquad \l \geq 0 \nonumber \\
\sgn (\lambda) &= -1  \, \qquad &\textrm{for} \qquad \l  < 0
\eea

This infinite sum  can be regularized in many ways. A natural regularization that arises from the path integral derivation is \eqref{etabeta}
\be\label{regpath}
\widehat\eta(\beta) := \sum_{\lambda} \sgn (\lambda)\,   \erfc{\left(|\l|\sqrt{\b}\right)} \, .
\ee
Another regularization  used in the original APS paper  \cite{Atiyah:1975jf}
is the $\zeta$-function regularization
\be\label{etas}
\etaaps (s)=\sum_{\lambda}\frac{\lambda }{|\lambda|^{s+1}}	
=\sum_{\lambda}\frac{\sgn (\lambda) }{|\lambda|^{s}}	\, .
\ee
The two regularization schhemes are related by  a Mellin transform 
\be
\etaaps(s) = \frac{s\sqrt{\pi}}{\Gamma (\frac{s+1}{2})}\int_{0}^{\infty} d\b \b^{\frac{s}{2}-1}\widehat \eta (\b) \, . 
\ee 
One can prove \cite{Atiyah:1976jg} that $\etaaps(s)$ is analytic near  $s =0$, so the $\eta$-invariant is given by
\be
\eta 
= \lim_{s \rightarrow 0 }\,  \etaaps(s) \, .
\ee
It is expected that the answer  is independent of the regularization up to local counter-terms that are implicit in the definition of a path integral. 

The  factor of half in front of $\eta$ in \eqref{aps} has the following consequence.  As one varies the metric on $\cN$, the eigenvalues of $\cB$ can pass through a zero and  $\eta $ would change by $\pm 2$.  The index then  changes by $\mp 1$ as expected for an integer. It also shows that neither the index nor the $\eta$-invariant are strictly topological and can change under smooth deformations. They  are nevertheless semi-topological in the sense they change only if the asymptotic data near the boundary is changed so as to alter the spectrum of $\cB$  to cause level crossing. 

\subsection{Elliptic genera and Jacobi forms \label{Elliptic}}

If the target space $\mathcal{M}$   is K\"ahler, then the worldsheet action in \eqref{sheet} has $(2, 2)$ supersymmetry. We assume that the theory is Weyl-invariant.  The nonlinear sigma model  then defines a $(2, 2)$ superconformal field theory  with left-moving and  right-moving super Virasoro algebras.  The elliptic genus of  an SCFT is  defined by
\begin{equation}
\chi (\t, z) =\TrH[] (-1)^{\tilde J + J} {e}^{2 \pi i \tau (L_{0} -\frac{c}{24})} e^{-2 \pi i \bar{\tau}( \tilde L_{0}-\frac{\tilde c}{24})} e^{2 \pi i z J} \, .
\end{equation}
where  $\ L_{0}$ and  $\tilde L_{0}$ are the left and right-moving Virasoro generators respectively, $c$ and $\tilde c$ are the central charges,  and  $J $ and $\tilde J$ are the  R-symmetry generators.  In our normalization 
$H= L_{0} + \tilde L_{0} -\frac{(c + \tilde c )}{24}$ generates time translations on the cylindrical worldsheet,  $P = L_{0}- \tilde L_{0}$ generates the space translations,  and $\tilde J$ can be identified with the right-moving fermion number. 
Since $J$ commutes with the right-moving fermion number $\tilde J$, the  elliptic genus can be thought of as a  right-moving Witten index.  Writing $\tau = \t_{1} + i \t_{2}$, one can identity $\beta = 2\pi \t_{2}$ and think of 
 $ 2\pi  \tau_{1} $ and  $2\pi z$ as the chemical potentials for the operators $P$ and $J$ respectively. 
 In the  path integral representation,  the presence of additional insertions has the effect of  twisting the boundary conditions along the time direction for the fields charged under $P$ and $J$. 

For a compact SCFT with central charge $c$,  the elliptic genus is a weak Jacobi form of weight $w=0$ and index $m= c/6$. 
Recall that  a Jacobi form\footnote{Our convention differs from \cite{Dabholkar:2012nd} where $k$ was used for weight and $m$ for index. We use $w$ for the weight here because in \S\ref{Cigar} for the cigar coset, $k$  corresponds to the level of the $SL(2,  \bR)$ WZW model.} $\phi(\tau, z)$ of weight $w$ and index $m$ is a  holomorphic function of  both $\tau$ and $z$ which is `modular in $\tau$' and `elliptic in $z$'. Thus, under modular transformation it transforms as
\begin{equation}
\phi\left(\frac{a \tau + b}{c \tau + d},\frac{z}{c\tau + d}\right)= (c\tau + d)^w\,  e^{\frac{2 \pi i m c z^2}{c \tau + d}}\phi(\tau,z) \, ;  \qquad 
\left(
\begin{array}{cc}a & b \\c & d\end{array}
\right) \in SL(2, \bZ)
\end{equation}
and under translations of $z$ it transforms as
\begin{eqnarray}
 \phi\left(\tau,z +\lambda \tau + \mu\right)=  e^{- 2 \pi \iimg m (\lambda^2 \tau + 2 \lambda z)}\phi(\tau,z)    \qquad with \qquad
\l, \m \in \bZ
\end{eqnarray}
A Jacobi form can be expanded in terms of $\vartheta$-functions i.e.
\be
    \phi(\tau, z) = \sum_{l \, \in  \mathbb{Z}/2m\mathbb{Z}}h_l (\tau) \vartheta_{m,l}(\tau,z)
\ee
where $h_l (\tau)$ is a modular form and $\vartheta_{m,l}(\tau,z)$ is  the index $m$ theta  function
\begin{eqnarray}\label{jacobi-theta}
    \vartheta_{m,l} (\tau,z)  := \sum_{n \, \in  \mathbb{Z}} q^{(l + 2mn)^2/4m}y^{l + 2mn} \, \qquad \qquad (q:=e^{2\pi i \t}\, , y := e^{2\pi i z}) \,  .
\end{eqnarray}
The modular invariance of the elliptic genus follows from its path integral representation. The path integral is  diffeomorphism invariant when regulated covariantly using a  covariant regulator such as a short proper time cutoff. It is also Weyl invariant for a  conformal field theory on the flat worldsheet\footnote{We consider nonchiral theories so  there is no possibility of gravitational anomalies.}. Consequently it is invariant under the mapping class  group $SL(2, \bZ)$ which is the group of global diffeomorphisms of the torus worldsheet  modulo Weyl transformations. Similarly, the elliptic transformation properties of the elliptic genus follow from the spectral flow \cite{Schwimmer:1986mf} of the left-moving superconformal field theory, and 
the theta expansion can  be understood \cite{Moore:2004fg}  physically by bosonizing the $U(1)$ R-symmetry current $J$.

For a compact SCFT,  the spectrum of $\tilde L_{0}$ is discrete and is paired by supersymmetry. Hence, only the right-moving ground states contribute to the elliptic genus and the elliptic genus is independent of $\bar \t$. 
This is essentially the same argument we used to show that the Witten index is independent of $\b$.  The holomorphic elliptic genus thus counts right-moving ground states with arbitrary left-moving oscillators.

For  a  non-compact target space, this argument fails. Therefore,  the noncompact elliptic genus need not be holomorphic. However, it is clear from its path integral representation that it must nevertheless have modular and elliptic transformation properties of a Jacobi form. As we explain in \S\ref{Mock} it  is given instead by the completion of mock Jacobi form. 

\section{Witten index and the $\eta$-invariant \label{Witteneta}}

At zero temperature,  only the ground states contribute to the Witten index. Thus, by definition,   the index $\cI$ of the Dirac operator equals $W(\infty)$:
\be
\cI := W(\infty)  \, . 
\ee
For a compact manifold  $\cM$ without boundary  or with a boundary and APS boundary conditions,  the Dirac Hamiltonian is self-adjoint. It has  a discrete  spectrum 
and its eigenvectors span the Hilbert space $\cH$. As a result, the Witten index is independent of $\beta$ and, in particular, 
\be
W(\infty)  = W(0) \, . 
\ee
This equality is the essential step in the proofs of both  the Atiyah-Singer and the Atiyah-Patodi-Singer index theorems because one can then evaluate the Witten index in the much simpler $\beta \rightarrow 0$ limit using the high temperature expansion of heat kernels.  

For  a compact target space without a boundary, the  index of a supersymmetric quantum mechanics \eqref{witten} has a path integral representation,   which is  the starting point  to obtain a derivation of the AS theorem using localization.  One would like  to similarly apply localization when $\cM$ has a boundary, but there is an obvious difficulty.   In general,   path integral formulation is much more subtle for a target space with a boundary because one cannot use the canonical measure.   For this reason,  it is convenient to map the problem to computation of the Witten index  $\hW$  of a noncompact manifold $\hM$ without boundary. It will also lead to a `spectral theoretic' reformulation of the APS theorem. \\

\subsection{Noncompact Witten  index \label{Noncompact}}

The APS boundary conditions imply  that for every solution of the Dirac Hamiltonian $H$ on $\cM$, there is an $L_{2}$-normalizable solution of the extended Dirac Hamiltonian $\widehat H$  on $\hM$.  One can therefore aim to express the Dirac index in terms of the noncompact Witten index $\widehat W(\infty)$ at zero temperature, which admits a simpler path integral representation. 
One immediate problem with this idea is that  the   spectrum  of  $\widehat H$ is expected to contain delta-function normalizable scattering states with a continuous spectrum in addition to the  $L_{2}$-normalizable states. 
It is not clear then that the operator $(-1)^{F} e^{-\beta \widehat H}$  is `trace class' in the conventional sense because  it may not have a  convergent trace after including the scattering states.  Thus, even before trying to develop the path integral for $\hW(\b)$,  it is necessary to first give a proper definition for it in the canonical formulation that correctly generalizes \eqref{witten}.

A natural formalism for this purpose is provided by  `rigged Hilbert space' or  `Gel'fand triplet' which generalizes the Von Neumann formulation of quantum mechanics based on a Hilbert space \cite{gelfand,Bohm:1989}. An advantage of this formalism  is that one can discuss spectral theory of  operators with  a continuous spectrum with `generalized' eigenvectors which may not be square-integrable. We  review some of the relevant concepts as they apply in the present context.  

The first Von Neumann axiom states that every physical system is represented (up to a phase) by a  vector in a Hilbert space $\cH$ with unit norm. This is essential for the Born interpretation because  the total probability of 
outcomes of measurements for  any physical system must be unity. The second axiom requires  that every physical observable corresponds to a self-adjoint operator on $\cH$.  This, however, is not always possible. A simple counterexample  is a free particle on a line $\bR$ with the Hamiltonian $H = p^{2}$. The self-adjoint operator corresponding to $H$
on $\cH$ has no normalizable eigenvectors,  so the set of eigenvalues of this operator is empty. On the other hand,   on physical grounds  one expects the free particle to have continuous energy  with  a sensible classical limit.   To deal with such more general physical situations,  it is necessary to   relax the second axiom and represent physical observables by operators defined on a domain in a  rigged Hilbert space using a Gel'fand triplet rather than  on a domain in  a Hilbert space. 

For  a quantum particle on a real line,  the Gel'fand triplet consists of  a Hilbert space $\cH$, the Schwartz space $\cS$, and the conjugate Schwartz space  $\cS^{\times}$. 
The Hilbert space  $\cH$ is  isomorphic to the  space $L_{2}(dx, \bR)$ of square-integrable  wave  functions  on 
$\bR$:
\be
\cH = \{ |\psi\rangle  \} \qquad \textrm{with} \qquad \langle \psi | \psi \rangle := \int dx\, \psi^{*}(x) \psi(x) < \infty \, ;
\ee
The Schwartz space  is the space of  infinitely differentiable `test functions' with exponential fall off. The conjugate 
Schwartz space $ \cS^{\times}$ is the set  $\{ | \phi\rangle\}$ such that 
\be
| \phi\rangle \in  \cS^{\times} \qquad \textrm{if} \qquad \langle \psi | \phi \rangle < \infty \qquad \forall  \quad | \psi\rangle \in  \cS \, .
\ee
The Gel'fand triplet provides a rigorous way to define the bra and ket formulation of Dirac and offers a way to discuss the spectral theory of  operators with continuous eigenvalues \cite{gelfand,Bohm:1989}. The notion of the Schwartz space is motivated by the fact that it is left invariant by unbounded operators like the position operator $x$. The conjugate Schwartz space  $ \cS^{\times}$ is where objects like the Dirac delta distribution $\delta(x)$ and plane waves $e^{ipx}$ reside. The elements of $\cS^{\times}$ need not have finite inner product with themselves and hence may not  be  
square-integrable,  but they have finite overlap with `test functions' belonging to $\cS$.  

Consider now a self-adjoint Hamiltonian $H$ defined on a domain $\mathcal{S} \subset \cH$. One can define the conjugate Hamiltonian $H^{\times}$ acting on $| \phi \rangle \in 
\cS^{\times}$ by the equation
\be
\langle \psi | H^{\times}| \phi \rangle = \langle H \psi | \phi \rangle\qquad \forall \quad |\psi \rangle \in \cS
\ee
With this definition, the eigenvalue equation for $H^{\times}$ 
\be
H^{\times}  | E \rangle = E  | E \rangle \, ,  \qquad  | E \rangle \in \cS^{\times}
\ee
should be  interpreted in terms of the overlap with test functions:
\be
\langle \psi  | H^{\times}  | E \rangle = E \langle \psi  | E \rangle \qquad \forall \quad |\psi \rangle \in \cS \, .
\ee
A `generalized eigenvector'  $| E \rangle$    may  lie outside the Hilbert space $\cH$ and may not  be normalizable. This means that it  cannot be prepared in any experimental setup. Nevertheless, the set $\{ | E \rangle \}$   provides a complete basis  in the sense that any state in $\cH$ can be expanded in terms of 
$ \{| E \rangle \}$. This is the content of the  Gel'fand-Maurin spectral theorem \cite{gelfand, maurin}.

For the example of a free particle discussed earlier, the operator $H^{\times}$  has the same formal expression as  $H$ as a differential operator:
\be
H^{\times} = -\frac{d^{2}}{dx^{2}} \, .
\ee
However, the  domain $D(H^{\times})$ is much larger than the domain $D(H)$.  This extension of the Hamiltonian  is  diagonalizable in the larger space 
$\cS^{\times}$ with generalized eigenfunctions $ \{e^{i p x}\} $ and eigenvalues $\{p^{2}\}$. 
In any lab with finite extent, one can  never experimentally realize an exact plane wave but only a wave packet that is sufficiently close to the energy eigenfunction.
Nevertheless, the plane waves form a complete basis in the sense that a square-integrable function in $\cH$ can be Fourier-expanded in terms of plane waves.
We denote the total space of generalized eigenvectors of $H^{\times}$ by $Sp$ which may contain  both the square-integrable bound states with discrete energies as well as  nonnormalizable scattering states with continuous energies.

Normally, one can gloss over these niceties essentially because of locality. A particle on an infinite line is an extreme idealization in a universe which may be finite. One expects  that  measurements of local quantities such as  scattering cross-sections in a particle physics experiment in a lab should not be affected by boundary conditions imposed at the end of the universe.  One should  arrive at the same physical conclusions whether one uses periodic or Dirichlet boundary conditions in a large box,  as one indeed finds in textbook computations. 

In the present situation, we are interested in global properties that depend sensitively on the boundary conditions. For example,  one cannot impose Dirichlet boundary  condition while preserving supersymmetry. The Gel'fand triplet  provides an appropriate  formulation so that one can discuss the scattering states without the need to put any boundary conditions to `compactify' space.
With these preliminaries, one can define the noncompact Witten index by
\be\label{witten2}
\hW(\b) := \TrS \left[  (-1)^{F} e^{-\beta \widehat H} \right]
\ee 
On the face of it, this definition is still not completely satisfactory. Even though the spectrum $Sp$ over which one traces now has a precise meaning, 
it is not clear that the trace thus defined  actually converges. 
For example, for a free particle, the heat kernel is well-defined\footnote{In what follows, we will use  $H$ instead of  $H^{\times}$ when there is no ambiguity to unclutter the notation.}
\be
K(x, y; \b)  = \langle x | e^{-\beta H} | y \rangle \, = 
\frac{1}{\sqrt{4\pi \b}}\exp\left[{-\frac{(x - y)^{2}}{4\b}}\right] .
\ee
However, if we try to define a trace then there is the usual `volume' divergence:
\be
\int dx \langle x | e^{-\beta H} | x \rangle = \frac{1}{\sqrt{4\pi \b}} \int dx \rightarrow \infty \, .
\ee
One might worry that the noncompact Witten index is also similarly divergent. 
Fortunately, the Witten index is a supertrace or equivalently a trace over  the \textit{difference} between two heat kernels corresponding to the bosonic and fermionic Hamiltonians $H_{+}$ and $H_{-}$ respectively.  If there is a gap between the ground states and the scattering states, then the two Hamiltonians differ from each other only over a region with compact support in  $\bR$. As a result,  the volume divergent contribution cancels in the supertrace. In the path integral representation,  this corresponds to the fact that the supertrace involves integrals over the `fermionic zero modes' in addition to the `bosonic zero mode' $x$. Under suitable conditions, the fermionic Berezin integration localizes the bosonic integral to a compact region on the real line  to yield a finite answer. In particular, the path integral receives vanishing contribution from the asymptotic infinity in field space. We elaborate on this point   in \S\ref{Worldpoint} and use it  for applying localization using \eqref{locdeform}.

Within this framework, one can now express the index of the Dirac operator on the original manifold $\cM$ with a boundary in terms of a  noncompact Witten index $\hW(\beta)$ on $\widehat\cM$. Assuming that the continuum  states in $Sp$ are separated from the ground states by a gap, at zero temperature only the $L_{2}$-normalizable ground states contribute to $\hW(\infty)$. Since these states are in one-one correspondence with the ground states in the original Hilbert space $\cH$ on $\cM$ with APS boundary conditions, we conclude
\be\label{index2}
\cI = \hW(\infty) \, .
\ee
In the limit of $\b \rightarrow 0$,  one can evaluate the Witten index by using the short proper time expansion of the heat kernels to obtain a local expression. It must correspond to the Atiyah-Singer term but now evaluated over $\hM$:
\be\label{as}
 \hW(0) = \int_{\hM} \alpha = \int_{\cM} \alpha
\ee
 where the second equality follows from the fact that the topological index density vanishes on  the half-cylinder $\cN \times \bR^{+} $. We can therefore write
\be\label{apsw}
\cI = \widehat W(0) +  ( \widehat W(\infty) -  \widehat W(0)) \, .
\ee
The  term in the parantheses is no longer zero and is in fact related to the $\eta$-invariant as will show in  \S\ref{Scattering}.   It is convenient to consider a regularized quantity
\be\label{etabeta}
 \widehat\eta (\b) := 2( \widehat W(\b) -  \widehat W(\infty))
\ee
which in the limit $\b \rightarrow 0$ reduces to the bracket in \eqref{apsw}. 
This provides a natural regularization described earlier in \eqref{regpath}.
With these identifications, equation \eqref{apsw} can be viewed as the statement of the APS theorem; the discussion above together with \S\ref{Scattering} can  be viewed as a rederivation of the APS result.
The noncompact Witten index is in general   $\b$-dependent because at finite temperature the  scattering states also  contribute. The bosonic and fermionic density of states in this continuum may not be exactly equal and need not cancel precisely.  The $\eta$-invariant of the boundary manifold $\cN$ thus measures the failure of the Witten index of the noncompact manifold $\hW$ to be temperature independent\footnote{This was noticed earlier in \cite{Alves:1986tp} in a special example.}.

\subsection{Scattering theory and the APS theorem\label{Scattering}}

There is a simpler way to  compute  $\hW(0)$ that makes this connection with the bulk Atiyah-Singer term \eqref{as} more manifest and easier to relate it to a path integral.  One can simply double the manifold to $\overline \cM$ by gluing its copy as in Figure \textbf{\ref{fig3}} as was suggested in \cite{Atiyah:1975jf}.  
\begin{figure}[htbp] 
\begin{center}
\begin{tikzpicture}[line width=1.5 pt, scale=.3]
\draw [blue] (-10,-1) .. 
controls (-3,-4) and (3,-4) .. 
(10,-1);
\draw  [blue] (-10,-9) .. 
controls (-3,-6) and (3,-6) .. 
(10,-9);
\draw  [blue] (-10,-1) .. 
controls (-20,2) and (-20,-12) .. 
(-10,-9);
\draw  [blue] (10,-1) .. 
controls (20,2) and (20,-12) .. 
(10,-9);
\node [genus, blue, draw, ultra thick, scale=1.5] at (-11, -5) {};
\node [genus, blue, draw, ultra thick, scale=1.5] at (11, -5) {};
\begin{scope}[shift={(0,-5)}, scale=2.5]	   
\draw [red] (0,0) ellipse (15 pt and 20pt);
\end{scope}
\node[] at (-21,-5) {$\overline{\cM} $}; 
\end{tikzpicture}
\end{center}
\caption{The doubled compact manifold $\overline{\cM}$ without boundary.}\label{fig3}
\end{figure}
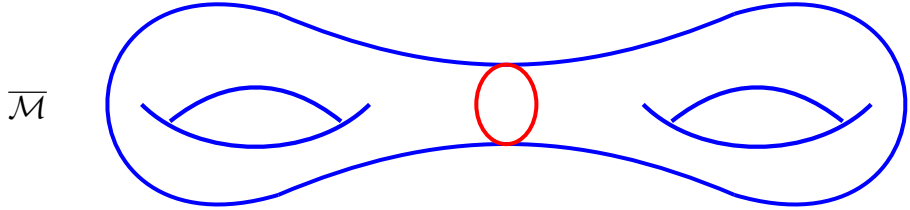
Since $\overline \cM$ is a manifold without a boundary, there is no contribution from the $\eta$-invariant. Moreover, by the reasoning 
before \eqref{as} the $\b \rightarrow 0$ expansion is local and gives the  Atiyah-Singer index density. 
In summary,
\be
\hW(0) = \half \overline W(0) = \int_{\cM} \alpha
\ee

To prove the APS theorem, we would like to show that the term in the parantheses in \eqref{apsw} equals the $\eta$-invariant. 
We note that the spectrum $Sp(H)$ of the Hamiltonian on $\hM$ is a direct sum of the discrete spectrum  of bound states $Sp_{b}(H)$   and  the continuum spectrum of  scattering states $Sp_{s}(H)$. Therefore the Witten index admits a spectral decomposition
\be
\hW(\b) = \TrSb \left[  (-1)^{F} e^{-\beta \widehat H} \right] +  \TrSs \left[  (-1)^{F} e^{-\beta \widehat H} \right] 
\ee
Since the continuum of states is separated from the zero energy states, it is clear that the first term can be identified with the index 
$\cI$ and hence with $\hW(\infty)$. To prove \eqref{etabeta}, we thus need to show that the contribution from the continuum equals the $\eta$-invariant:
\be\label{etaspec}
2\TrSs \left[  (-1)^{F} e^{-\beta \widehat H} \right] = \widehat \eta(\b) \, .
\ee
We show this by relating  the supertrace to  the difference in the density of bosonic and fermionic scattering states\footnote{The relation between $\eta$ invariant and scattering theory was observed earlier in special cases in \cite{Akhoury:1984pt, Niemi:1985ht, Forgacs:1987ti}.} on $\widehat \cM$ which in turn can be related to the difference in phase shifts.  

Asymptotically, the metric on $\hM$ has the form \eqref{productmetric} with $0 < u < \infty$. We can use separation of variables to first diagonalize the operator $\cB$ on $\cN$ with eigenvalues $\{ \l \}$. The Dirac operator on manifold $\widehat \cM$ can be expressed in terms of eigenvalues $\lambda$ of the boundary operator $\cB$ as in \eqref{dirac1}.  
The asymptotic form of the scattering wave functions is then
\begin{eqnarray}
	\psi_+^{\l k}(u)&\sim& c_+^\l\left[ e^{\iimg ku}+e^{\iimg \delta^{\l}_+(k)-\iimg ku}\right] 
\nonumber	
\\
	\psi_-^{\l k}(u)&\sim& c_-^\lambda\left[ e^{\iimg ku}+e^{\iimg \delta^{\l}_-(k)-\iimg ku}\right] 	
\label{spectraltheory9}		
\end{eqnarray}
where $\delta^{\l}_{\pm}(k) $ are the phase shifts. 
The trace  \eqref{etaspec} over scattering states  can be expressed as 
\begin{eqnarray}
	 2 \sum_{\l}\int  dk\,  \Big[\rho^{\l}_+(k)-\rho^{\l}_-(k)\Big]e^{-\beta E(k)}
\label{spectraltheory7}		
\end{eqnarray}
where $\rho^{\l}_+(k)$ and $\rho^{\l}_-(k)$ are the density of bosonic  and fermionic states of the theory for the $\l$ subsector. 
Using a standard result \eqref{spectraltheory21} from  scattering theory, which we review in \S\ref{Spectral}, we can relate the difference in the density of states to the difference in phase shifts
\be
 \rho^\l_+(k) -\rho^\l_-(k) =  \frac{1}{\pi}  \frac{d}{dk} \left[\delta^{\l}_+(k) -\delta^{\l}_-(k)\right] \, .
 \ee
 In general the individual phase shifts and density of states are nontrivial functions of $k$ that depend on the details of the manifold $\hM$. After all,  they contain  all the information about the S-matrix. The exact form of the scattering states similarly has a complicated functional dependence on $u$. Generically, it would be impossible to compute any of them exactly.  Remarkably,  the \textit{difference} between the phase shifts is determined entirely by the  asymptotic data as a consequence of supersymmetry relation \eqref{apsindexth72a} in the asymptotic region.
By substituting the asymptotic wave-functions \eqref{spectraltheory9} into  \eqref{apsindexth72a} we obtain

\begin{eqnarray}
c_{+}^{\l}\sqrt{E}\Big[ e^{\iimg k   u} +e^{\iimg \delta^{\l}_{+}} \,  e^{-\iimg k   u}\Big]
&=&
c_{-}^{\l}\Big[(-\iimg k  + \l ) e^{\iimg k   u} +e^{\iimg \delta^{\l}_{-}} \, (\iimg k  + \l  )  e^{-\iimg k   u}\Big]
\end{eqnarray}
with $E = k^{2 } + \l^{2}$.  This implies
\be
\frac{c_+^\l}{c_-^{\l}}= \frac{(-\iimg k  + \l  )}{\sqrt{E}} \, , \qquad
\frac{e^{\iimg \delta_{+}^{\l}} }{e^{\iimg \delta_{-}^{\l}} }
=	-\frac{(\iimg k +\l  )}{(\iimg k  -\l  )} 
\ee
and therefore, 
\begin{eqnarray}
	\delta^{\l}_+(k) -\delta^{\l}_-(k) = -\iimg\,   \ln \, \left( \frac{\iimg k+\lambda}{\iimg k-\lambda }\right)+\pi 
\label{spectraltheory10}		
\end{eqnarray}
in each eigensubspace with eigenvalue $\lambda$. 

Now we use the formula \eqref{spectraltheory10} in \eqref{spectraltheory21} to get 
\begin{eqnarray}
 \rho^\l_+(k) -\rho^\l_-(k)
=-\frac{2\lambda}{\pi(k^2+\lambda^2)}
\label{spectraltheory22}		
\end{eqnarray}
After summing over all $\l$ we obtain
\begin{eqnarray}
2\TrSs \left[  (-1)^{F} e^{-\beta \widehat H} \right] 
 &=&	\sum_{\l}\int_0^\infty \frac{dk}{2\pi }\,  \left[ \rho^\l_+(k) -\rho^\l_-(k) \right]\, e^{-\beta(k^2+\lambda^2) } \nonumber\\
 &=& \sum_{\l}\sgn(\lambda)\, \erfc \left(|\lambda |\sqrt{\frac{\beta}{2}}\right)
\label{spectraltheory23}		
\end{eqnarray}
This is precisely the regulated expression \eqref{etabeta} for the $\eta$-invariant of the boundary operator. 
We have thus proven 
\be
\cI = \widehat W(0) +  ( \widehat W(\infty) -  \widehat W(0))  = \int_{\cM} \, \a -\half \eta
\ee
which is the Atiyah-Patodi-Singer index theorem. 

\section{The $\eta$-invariant and path integrals \label{Path}}

Given the definition of the noncompact Witten index in \S\ref{Noncompact} one can use its path-integral representation and use localization methods to compute it. In \S\ref{Cigar} we show how this works for the  two-dimensional  surface of a finite cigar with a boundary by relating its index to the Witten index of the infinite cigar. In this simple example, one can explicitly evaluate the Witten index using localization and compare with the $\eta$-invariant obtained from operator methods. 

Can one  find a way to formulate a path integral that directly computes the $\eta$-invariant without the bulk Atiyah-Singer piece? 
This can be achieved as follows. As we have observed in \S\ref{Scattering}, given a manifold with boundary $\cM$ such that metric is of the product form near the boundary we can {\it trivially} extend the manifold to a non-compact manifold $\mathcal{\widehat M} $. In $\mathcal{\widehat M} $,  the $\eta$-invariant gets contribution only from the scattering states of $\widehat \cM$. Now the scattering states of $\mathcal{\widehat M} $ are the same as the scattering states of $\mathbb R^{+} \times \cN$ with APS boundary condition at the origin. We use this physical picture to find a path integral representation to compute $\eta$ invariant. First we will explain the space-time picture and then we will map it to a world-line computation. 

The Dirac operator on the half line is given by $      \widetilde{ \slashed{ D} } =     \g^{u}(\partial_u\ +  \bar\gamma \cB )$. We can diagonalize the boundary operator as in \eqref{apsindexth71a}, \eqref{apsindexth71b}.  Effectively, for each eigenvalue $\l$ of the boundary operator $\cB$, we have a supersymmetric quantum mechanics on a half line. The APS boundary condition is essentially Dirichlet boundary condition for one chirality and Robin boundary condition for the other chirality. To obtain a path integral representation with these boundary conditions,  it is more  convenient to `double' the manifold $\mathbb R^{+} \times \cN$ to obtain a noncompact cylinder $\widetilde \cM := \mathbb R \times \cN $ (see Figure \textbf{\ref{fig4}}) without any boundary. We extend it in a manner that is consistent with the APS boundary conditions.  The manifold  $\widetilde \cM := \mathbb R \times \cN $ has parity symmetry 
\be
\label{paritytransform}
	P: \qquad \qquad u\rightarrow -u \qquad,\qquad \psi_{\pm }\rightarrow -\psi_{\pm }
\ee
that is consistent with supersymmetry and leaves the supercharge invariant. 
The path integral on the original manifold  $\mathbb R^{+} \times \cN$ with APS boundary conditions can be obtained  by considering the path integral on the manifold $\widetilde \cM := \mathbb R \times \cN $ projected onto $P$ invariant states. This is effected by  the insertion of the following operator 
\begin{eqnarray}
\frac{1}{2}\Big[1+P\Big] \,
\label{projectionoperator1} 	
\end{eqnarray} 
where $P$ is the parity operator. Invariance under the reflection  of $u$ keeps only parity-even wave functions  in the trace for one chirality, effectively imposing Dirichlet boundary condition on the half line. Supersymmetry ensures that the other chirality satisfies the Robin boundary condition as required by the APS boundary conditions. See,  for example,  \cite{Farhi:1989jz, Troost:2017fpk} for a more detailed discussion.

Since we are interested in the operator $   \widetilde{ \slashed{ D} } =   \g^{u}(\partial_u\ +  \bar\gamma \cB )$ on the half-line, the extension of this operator should transform as an eigen-operator under parity. Given $u$ transform as in $\eqref{paritytransform}$ we are left with the choice 
\be
 \cB \rightarrow -\cB .
\ee
This ensures that Dirac operator as a whole is invariant. 
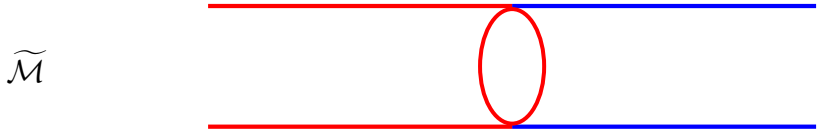
\begin{figure}[htbp] 
	\begin{center}
	        \vskip 9mm
		\begin{tikzpicture}[line width=1.5 pt, scale=.8]
		\draw[blue, ultra thick] (0,-1)-- (5,-1)  ;
		\draw[blue, ultra thick] (0,1)-- (5,1)  ;
		\draw [red] (0,0) ellipse (15 pt and 27pt);
		\draw[red, ultra thick] (0,-1)-- (-5,-1)  ;
		\draw[red, ultra thick] (0,1)-- (-5,1)  ;
		\node[] at (-8, 0) {$\mathcal{\widetilde M} $}; 
		\end{tikzpicture}
		\vskip 3mm
	\end{center}
	\caption{The doubled noncompact cylinder $\widetilde \cM = \cN \times \bR$. }\label{fig4}
\end{figure}
The extended Dirac operator on the doubled cylinder thus takes  the form 
\be
    \widetilde{ \slashed{ D} } =   \g^{u}(\partial_u\ + \varepsilon(u) \bar\gamma \cB )
\ee
instead of \eqref{apsdirac} where $\varepsilon(u)$ is a step function with a discontinuity at $u=0$.  One can also take $\varepsilon(u)$ to be a smooth smearing function which interpolates between $-1$ to $1$ as $u$ varies from $-\infty $ to $+\infty$ to obtain a smooth Dirac operator. One example of such function is $\tanh(u)$.  This does not change the conclusions because the $\eta$-invariant does not change under deformations that do not change the asymptotics.

Now we return to the world-line picture. For each eigenvalue of $\cB$ we can map this problem to a world-line path integral problem. The effect of the eigenvalue of the boundary operator can be incorporated by adding a super-potential $h$. It  reduces the problem to computing Witten index $\widetilde{W}(\beta)$ in the presence of  a super-potential with a target space being half-line appropriately extended to the full line.  The Witten index can be computed using path integral.   The corresponding supersymmetric quantum mechanics will have a superpotential $h(u) = \varepsilon(u) \l$ determined by the eigenvalue $\l$. With this construction, we conclude
\be\label{etatilde}
\widehat \eta(\b) = \widetilde \eta (\beta) = 2 ( \widetilde W(\beta) - \widetilde W(\infty) )
\ee
It is straightforward to write a path integral representation for  $\widetilde \eta(\b)$ on the  non-compact cylinder $\widetilde \cM$ which is much simpler than the path integral on $\hM$. 
In \S\ref{Callias}, we compute it using localization and relate it to Callias index \cite{Callias:1977kg, Bott:1978bw}.

\subsection{Supersymmetric worldpoint integral \label{Worldpoint}}

Some of the essential points about a noncompact path integral can be illustrated by a `worldpoint'  path integral where the   base space $\Sigma$ is a point and the  target space  $\cM$ is a the real line $ -\infty <  u  <\infty $. We discuss this example first before proceeding to localization. 
The supersymmetric worldpoint action is given by 
\begin{eqnarray}
	S(u, F, \psi_-,\psi_+) =    \frac{1}{2} F^{2} + \iimg F \,  h'(u)
	+ \iimg h''(u)\,  \psi_- \psi_+
\label{wpresult1}
\end{eqnarray}
where 
\be
h'(u) := \frac{dh}{du}\, , \qquad h''(u) = \frac{d^{ 2}h}{du^{2}} \, . 
\ee
The action 
can be obtained from the euclidean continuation\footnote{Note that in the Euclidean continuation $F \rightarrow \iimg F$. So,  the limit $\frac{\partial}{\partial \tau} = 0$ of the Euclidean  and the Lorentzian actions gives different actions for the supersymmetric integral.} of \eqref{susysigma5} by setting
\begin{eqnarray}
\frac{\partial}{\partial \tau} = \frac{\partial}{\partial \s} =0 \,,\qquad    g_{11}(u) =1\, .
\end{eqnarray}
The path integral is now just an ordinary superintegral with flat measure
\bea
W(\beta) = -\iimg \int_{-\infty}^{\infty} du \int_{-\infty}^{\infty} dF \int  d\psi_-\, d\psi_+ \, 
\exp\left[-\b S(U) \right] \, .
\eea
The  normalization factor $-\iimg$ can be understood as follows. For $2n$ real fermions we have 
\be
\bar\gamma = \iimg^{n} \gamma^{1} \dots\gamma^{2n} = (-2i )^{n} \psi^{1}_0\dots\psi_0^{2n}\qquad  \, .
\nonumber\ee
Moreover, $\Tr\,  \bar \gamma^{2} = 2^n$ which is the dimension of the spinor representation 
\be
\Tr\,  \bar \gamma^2 = N \int (-2i)^{n} \psi^{1}_0\dots\psi_0^{2n}  d\psi^{1}_0 \dots d\psi_0^{2n}
\nonumber
\ee
which implies $N= (-i)^{n}$. We have  two real fermions, hence $n=1$ and  $N= -i$. 

A particularly interesting  special case is
\be\label{htanh}
h'(u) = \l \tanh (a u) \, , 
\ee
for real $\l$. Integrating out the fermions and the auxiliary field $F$ gives
\bea
W(\beta) = -\sqrt{\frac{\beta}{2 \pi}}  \int_{-\infty}^\infty dx\, 
h''(u)\exp\left[- \frac{\beta}{2}(h'(u))^2 \right] 
\label{wpresult7}
\end{eqnarray}
 One can change variables 
\be
y = \sqrt{\frac{\b}{2}}h'(u)  \, , \quad  dy = \sqrt{\frac{\b}{2}}h''(u) dx
\ee
As $u$ goes from $-\infty$ to 
$\infty$, $y(u)$ is monotonically  increasing or decreasing  depending on if $\l$ is positive or negative; the inverse function $u(y)$ is single-valued,  and the integral  reduces to
\begin{eqnarray}
W(\beta) &=& -\frac{1}{\sqrt{\pi}} \int_{ -\sqrt{\frac{\b}{2}}\l}^{\sqrt{\frac{\b}{2}} \l} dy \, 
e^{-y^{2}} \, \nonumber \\
&=&  - \sgn (\l) \,  \erf \left(\sqrt{\frac{\b}{2}} |\l |\right) 
\label{wpresult8}
\end{eqnarray}

The  worldpoint integral illustrates a number of important points. 
\begin{myenumerate}
\item
Without the fermionic integrations, the integral has a volume divergence because $h'(u)$ is bounded above for large $|u|$.  Inclusion of fermions effectively limits the integrand to the region close to the origin where $h'(u)$ varies,  and makes the integral  finite. 
\item
In the limit $\l \rightarrow 0$,  the action reduces to that of a free superparticle. In this case,  the integral is of the form $\infty \times 0$ and is ill-defined. Regularizing with $\l$ yields different answers depending on whether we approach $0$ from positive or negative side. This is related to the jump in the $\eta$-invariant when an eigenvalue  of the boundary operator $\cB$ crosses a zero in a spectral flow as explained before figure \textbf{\ref{fig5}}. 
\item
The answer depends only the asymptotic behavior of $h'(u)$ at $\pm \infty$ and is independent of any deformations that do not change the asymptotics. In  particular,  one would obtain the same result in  the limit $a \rightarrow \infty$ in \eqref{htanh}, when $h'(u)$ can be expressed in terms of the Heaviside step function:
\be 
h'(u) = \l \Big[\theta (u) - \theta (-u)\Big] \, .
\ee
\item
The  error function \eqref{wpresult8} which appears naturally in this integral makes its appearance  in the proof of the APS theorem \cite{Atiyah:1975jf, Atiyah:1976jg} and also  in the  definition of the completion  \eqref{defstar} of a mock modular form and, in particular, in \eqref{gcompletion}. This is not a coincidence. The two turn out to be related through a path integral which localizes precisely to the ordinary superintegral considered above.  For this reason, this example is particulary important for  our discussions of  the  $\eta$-invariant  and its connection to mock modularity.
\end{myenumerate}
The worldpoint  integral does not have an operator interpretation in terms of a trace over a Hilbert space. To see the connection with the canonical formalism, we consider in \S\ref{Callias} the worldline version corresponding to the path integral for a supersymmetric quantum mechanics, and relate it to Callias-Seeley-Bott index theorem \cite{Callias:1977kg, Bott:1978bw} . After localization, the worldline path integral will reduce to the worldpoint integral considered above. 

\subsection{Callias index theorem and the $\eta$-invariant \label{Callias}}
 
In this section we compute the $\eta$ invariant by computing path intergal for supersymmetric quantum mechanics with target space $\widetilde \cM$. 
As discussed in \S\ref{Scattering} we can use separation of variables to first diagonalize the operator $\cB$ on $\cN$ with eigenvalues $\{ \l \}$ but now for the entire manifold  $\widetilde \cM$. For each eigenvalue $\l$, the problem reduces to a supersymmetric quantum mechanics with a one-dimensional target space and a superpotential $h(u)$. The path integral for this problem can be readily written down and has been considered earlier in \cite{Imbimbo:1983dg}. 
The action for SQM can be obtained as a specialization of \eqref{sheet} with target space as $\mathbb{R}$ by setting
\be
\frac{\partial}{\partial \s} =0 \, .
\ee
 The Euclidean action for the components of the superfield $U$ is 
\begin{eqnarray}
S[U, \b] &=&  \int_{0}^{\b}d\t \left[\frac{1}{2}\dot{u}^2 + \frac{1}{2}\psi_{-}\dot{\psi}_{-} +\frac{1}{2}\psi_{+}\dot{\psi}_{+} + \frac{1}{2} F^{2} + \iimg h'(u)  F + \iimg h''(u)\psi_{-}\psi_{+}  \right] 
\end{eqnarray}
with $h^\prime (u) = \l \tanh u$. To compute the $\eta$-invariant we need to have to evaluate the projected Witten index \eqref{projectionoperator1} 
\begin{eqnarray}
	\widetilde W(\beta) &=&\frac{1}{2}\tr\left[(-1)^F(1+P(-1)^F) \right] =\frac{1}{2}\tr\left[(-1)^F \right] +\frac{1}{2}\tr\left[P \right] 
\nonumber\\
&=&	\frac{1}{2}\widetilde W_1(\beta)+	\frac{1}{2}\widetilde W_2(\beta)
\label{hhspacewindexdefn1}
\end{eqnarray}
The path integral for the first term is the same as before with periodic boundary conditions for bosons and fermions.  In the path integral for the second term   both the bosons and the fermions have anti-periodic because of the insertion of $P(-1)^F$.

Given a  path integral representation for the Witten index \eqref{witten2}, one can apply localization methods by deforming it  by a $\scharge$-exact term:
\begin{eqnarray}
\widetilde W(\beta, \xi)=	\int [dX]\,  	e^{-  S[X, \b]-\xi \scharge V[X]} \, .
\label{eqn2.2}
\end{eqnarray}
The integral is over the supermanifold  $\{X(\s)\}$ of maps from $\Sigma$ to the supermanifold $s\cM$ with a $Q$-invariant supersymmetric measure. 
The functional $V$ is chosen such that $\scharge V    \geq 0$ (so that adding $\scharge V  $ to the action does not blow up the path integral) and $\scharge^2  V   = 0$. This implies
\be
\label{locdeform}
\frac{d}{d\xi}
\widetilde W(\beta, \xi) =	\int [dX] 	\, \scharge  V  [X]\, e^{-S[X, \b]-\xi \scharge V[X]} =	\int [dX] 	\, \scharge\big(  V  [X]\, e^{-S[X]-\xi \scharge  V   [X]}\big) 
\ee
The supercharge $Q$ is a fermionic vector field  on this supermanifold so one can use Stokes theorem to show that the above integral vanishes. Usually one use the compactness of the manifold $\cM$ to argue that there are no boundary terms. In the noncompact case, there can be a contribution from the asymptotics in the field space but it can still vanish by the arguments outlined in our discussion of the noncompact Witten index. Hence $W(\beta, \xi)$ is independent of $\xi$. This implies that one can perform the integral $W(\beta, \xi)$ for any value of $\xi$ and in particular for $\xi \rightarrow \infty$. In this limit, the functional integral localizes onto the critical points of the functional $\scharge V[X]$. 
A canonical choice of $V$ is
\begin{eqnarray}
V= \sum_i (\scharge \psi^i)^\dagger \psi_i  \, .
\end{eqnarray}
The critical points of this function are simply the $Q$-invariant supersymmetric configurations $\{ X_{0} (\mu) \}$ satisfying $\scharge \psi=0 $ and $(\scharge \psi)^\dagger=0$. The critical manifold is   parametrized the collective coordinates $\{\mu\}$.  The path integral localizes to an integral on the critical manifold 
 \begin{eqnarray}
\widetilde W(\beta) = \int _{\mathcal{X}}	[d\mu]\, e^{-S[X_0(\mu), \b]}	\frac{1}{\textrm{SDet} (\mu) }
\end{eqnarray}
where $\textrm{SDet}(\mu)$ is the superdeterminant  of the quadratic fluctuation operator  coming from   $QV$ action expanded around $X_{0}(\mu)$. See  \cite{Pestun:2016jze} for a recent review. 

For the problem at hand, we choose
\begin{eqnarray*}
V &=& \psi_{+} (\scharge_{+} \psi_{+}) =  \iimg \psi_{+} \dot u\\
\scharge_+ V & =& \dot u^2 + \psi_{+} \dot \psi_{+}
\end{eqnarray*}
where $\scharge_+$ is a real supercharge. 
The path integral localizes to constant modes of $u$ and $\psi_{+}$. Fluctuations around the constant modes  are given by 
\begin{eqnarray}
u = u_0 + \frac{1}{\sqrt{\xi}}\tilde u \qquad \psi_{+} = \psi_{+0} + \frac{1}{\sqrt{\xi}}\eta
\label{Localizationfluctuation1}
\end{eqnarray}
with $\tilde u $ and $\eta$ satisfying periodic boundary conditions. 
So we have,
\begin{eqnarray*}
\widetilde W_1[\beta] &=&- \iimg  \int du_0 [dF][d\tilde u] [d\psi_{-}] d\psi_{+0} [d \eta]\exp  \left(-S[X_{0}, \b] - \xi (\scharge_+ V)^{[2]} \right)
\end{eqnarray*}
Expanding the $\psi_{-}$ in modes and after evaluating the non-zero mode integrals we obtain
\be
\label{calliasint}
\widetilde W_1(\beta) =-
 \iimg \int  \frac{d u_0}{\sqrt{2 \pi \beta}}\, d\psi_{-0} \,d\psi_{+0} \exp \left(-\frac{\beta}{2} (h'(u_0))^2 - \iimg \beta (h''(u_{0})\psi_{-0}\psi_{+0}\right)
\ee
The factor of $\frac{1}{\sqrt{2 \pi \b}}$ comes from the determinants computed in  \S\ref{Ultra}.  The integral \eqref{calliasint} is identical to the worldpoint superintegral \eqref{wpresult7}. Hence we obtain
\begin{eqnarray}
\widetilde W_1(\beta)  = -\sgn (\l)\,  \erf \left(|\l|\sqrt{\frac{\beta}{2}}\right)
\end{eqnarray}
It remains to compute the other piece due to insertion of $P(-1)^F$. In this case, the path-integral localizes to $u=0=\psi_{+}$. Small fluctuations around the saddle point is given by  
\begin{eqnarray}
u = 0+ \frac{1}{\sqrt{\xi}}\bar u \qquad \psi_{+} =0 + \frac{1}{\sqrt{\xi}}\bar \eta
\label{Localizationfluctuation1}
\end{eqnarray}
Here $\bar u $ and $\bar \eta $ satisfy {\it anti}-periodic boundary condition.
\be
\widetilde W_2[\beta] =- \iimg\int    [dF][d\bar u] [d\psi_{-}]  [d \bar \eta]\exp  \left(-S[X_{0}, \b] - \xi (\scharge_+ V)^{[2]} \right) =1
\ee
So the full answer is given by \eqref{hhspacewindexdefn1}
\begin{eqnarray}\label{etaerror}
\widetilde W(\beta)=-\frac{1}{2}\sgn (\l)\,  \erf \left(|\l|\sqrt{\frac{\beta}{2}}\right)+\frac{1}{2}
\end{eqnarray}
We have performed the computation for a single eigenvalue $\lambda$. We get \eqref{etaerror} for each eigenvalue. From \eqref{etatilde} we get the $\eta$ invariant to be  
\be
\widetilde \eta(\beta) =\sum_\lambda  \sgn (\l)\left[1-\,  \erf \left(|\l|\sqrt{\frac{\beta}{2}}\right)\right]
=\sum_\lambda \sgn(\lambda) \erfc \left(|\l|\sqrt{\frac{\beta}{2}}\right)
\ee
which reproduces the expression \eqref{spectraltheory23} for $\widehat \eta(\b)$ obtained from scattering theory. Hence 
\be
\widetilde \eta(\beta) = \widehat \eta(\b)
\ee
In conclusion, the Witten index for the worldline quantum mechanics is temperature-dependent as a consequence of the noncompactness of the target manifold. 

\subsection{The $\eta$-invariant of a finite cigar \label{Cigar}}

It is instructive to apply the general considerations in earlier sections to  an explicit computation
for  a simple and illustrative example where $\cM$ is a two dimensional  \textit{finite} cigar with  metric
\begin{equation}
\label{cigarmetric}
ds^2 = k\, (dr^2 + \tanh ^2 r\,  d\theta^2) \, 
\end{equation}
where   $\theta$ is a  periodic with period $2\pi$ and $ 0 \leq r \leq r_{c}$. The manifold has a boundary 
at $r=r_{c}$  with a product form $\cN \times \bI$ where $\cN$ is the circle parametrized by $\theta$ with radius $\sqrt{k}$. 
The non-zero Christoffel symbols are
\begin{equation}
\Gamma_{r\theta \theta} = -\frac{1}{2}k\;\partial_r(\tanh^2r) \qquad \qquad \Gamma_{\theta\theta r} = \Gamma_{\theta r \theta } = \frac{1}{2 }k\;\partial_r(\tanh^2r)
\end{equation} \label{Cigar1}
The orthonormal forms and the nonzero vielbeins are
\bea\label{viel}
e^{1} &= \sqrt{k} dr \, \qquad e^{2} &= \sqrt{k} \tanh(r) d\theta \nonumber\\
e^{1}_{\, r} &= \sqrt{k}  \, \qquad e^{2}_{\, \theta} &= \sqrt{k} \tanh(r)   
\eea
The cigar has a Killing isometry  under translations of $\theta$ with the Killing vector 
\begin{eqnarray}
K^i = (0,1)  \, \qquad
K_i = g_{ij}K^j = (0, k \tanh^2 r) \, .
\end{eqnarray}\label{Cigar2}
The  $N = (0, 1)$ supersymmetric action can be obtained from  \eqref{action2} by setting
\begin{eqnarray}
    F^i = 0 \quad \quad \psi^i_- = 0 \quad \quad \psi_+ = \psi^i \, .
\end{eqnarray}
The Lorentzian action is given by:
\begin{equation}\label{e1}
I = \frac{1}{4 \pi}\int {d}^{2}\sigma \, {g}_{\tsa\tsb}\bigg( {\partial}_{\tau}{X}^{\tsa}{\partial}_{\tau}{X}^{\tsb} - {\partial}_{\sigma}{X}^{\tsa}{\partial}_{\sigma}{X}^{\tsb} + i {\psi}^{\tsa}{D}_{\tau - \sigma}{\psi}^{\tsb} \bigg) 
\end{equation}
We  dimensionally reduce along the worldsheet $\s$ direction to convert the action on the 2-torus to a collection of actions on a circle. 

Scherk-Schwarz reduction \cite{Scherk:1979zr} along the sigma direction using the Killing vector gives
\begin {eqnarray}
X^\tsa (\sigma + 2\pi) &=& X^\tsa(\sigma) + 2 \pi w K^\tsa 
\end {eqnarray}
where $w$ is the winding number. We have
\begin{equation}\label{e2}
\partial_\sigma X^\tsa = w K^\tsa \quad \textrm{and} \quad \partial_\sigma \psi^\tsa = -w \partial_\tsb K^\tsa \psi^\tsb
\end{equation}
where the derivative of $\psi^\tsa$ is deduced from the transformation of the superfield $X^\tsa = x^\tsa + \bar \theta \psi^\tsa + \bar\theta\theta F^\tsa $ under the Killing symmetry.
Using \eqref{e2} in action \eqref{e1} and integrating over the $\sigma$ direction we get the Euclidean action after a Wick rotation:

\begin{equation}
S= \frac{1}{2} \int d\t  \bigg( G_{ij}{\partial}_{\tau}{X}^{i}{\partial}_{\tau}{X}^{j} + {G}_{ij} w^2 K^i K^j  + {G}_{ij}{\psi}^{i}{D}_{\tau}{\psi}^{j} - \iimg w \psi^i {K}_{ij}\psi^j \bigg)
\label{cigarqm21}
\end{equation}
Plugging \eqref{cigarmetric} in the action \eqref{cigarqm21} we get: 
\begin{eqnarray}
S[\beta;k,w] &=& \int_0^\beta d\tau \frac{1}{2}\bigg( k\dot{r}^2 + k\tanh^2r \dot{\theta}^2 + w^2k \tanh^2 r + k\psi^r\dot{{\psi}^{r}} - k\psi^r \partial_r(\tanh^2r)\dot{\theta}\psi^\theta 
\nonumber\\
&&+  k\tanh^2r \psi^\theta\dot{\psi^\theta} -  \iimg w \psi^\theta \partial_r (k \tanh^2 r)\psi^r \bigg)
\label{cigarqm22}
\end{eqnarray}

Our goal is  to evaluate  the path integral on the  infinite cigar  using localization  and then connect it to the  $\eta$-invariant for a finite cigar. To deform the action we choose 
\begin{eqnarray}
V= G_{rr}\psi^r\delta\psi^r = k \psi^r \dot{r}
\label{cigarqm24}
\end{eqnarray}
This localizes the integral to constant modes of $r$ and $\psi^r$. We have:
\begin{eqnarray}
\hW (\beta) &=&- \iimg \int dr_0 d\psi^r_0[d \theta][d\psi^\theta ] \, \exp \bigg[-\int_0^\beta d\tau L (r_0,\psi^r_0,\theta, \psi^\theta) - \xi \int_0^\beta Q V^{[2]}\bigg] 
\label{cigarqm25}
\end{eqnarray}
To compute the quadratic fluctuations $Q V^{[2]}$ we set 
\begin{eqnarray}
r = r_0 +\frac{1}{\sqrt{\xi}}\chi
\quad \quad
\psi^r = \psi^r_0 +\frac{1}{\sqrt{\xi}}\eta^r
\label{cigarqm27}
\end{eqnarray}
so that the quadratic fluctuations are given by
\begin{equation}
\xi \int_0^\beta d\tau\,  Q V^{[2]} = \int_0^\beta d\tau ( k\dot{\chi}^2 +  k\eta^r\dot{{\eta}^{r}})
\label{cigarqm28} \, .
\end{equation}
The  transformation \eqref{cigarqm27} has  unit Jacobian. We can now mode expand $\theta$ and $\psi^\theta$ and we have
\begin{eqnarray}
\theta(\tau) &=& \frac{2 \pi p \tau}{\beta} + \sum_{m} \theta_m e^{2\pi \iimg m\tau/\beta}
\quad \quad
\psi^\theta(\tau) = \psi^\theta_0 + \sum_{m} \psi^\theta_m e^{2\pi \iimg m\tau/\beta}
\end{eqnarray}

After integrating out the fluctuations and non-zero modes of $\theta$ and $\psi^\theta$, we have
\begin{eqnarray}
\widehat W(\beta;k,w) =&& -\frac{ 2 \pi i }{2\pi \beta}   \int {dr_0}d\theta_0 d\psi^r_0 d\psi^\theta _0 \sum_p  \exp \bigg[-\int_0^\beta d\tau\,  S (r_0,\psi^r_0,\psi^\theta _0)\bigg] 
\nonumber\\
 =&&  -\frac{\iimg}{\beta}  \sum_p\int_0^{\infty} dr_0\frac{1}{2} k\partial_r ( \tanh^2 r)\Big|_{r_0}\left(- \iimg   w +\frac{2 \pi p}{\beta}\right) e^{-\frac{1}{2}\beta k\tanh^2 ( r_0) \left((\frac{2 \pi  p}{\beta})^2 +w^2\right)}
\label{cigarqm36}
\end{eqnarray}
The factor of $\frac{1}{{2\pi \beta}}$ from the determinants as before. 
Substituting
$
y = \frac{1}{2}\beta k\tanh^2r_0
$, we obtain
\bea
\widehat W (\beta;k,w) &=&-\frac{\iimg }{\beta} \sum_{p\neq 0}\int_0^{\frac{1}{2}\beta k} [dy] \left(- \iimg w +\frac{2 \pi p}{\beta}\right) \exp\left[-y \left(\left(\frac{2 \pi p}{\beta}\right)^2 + w^2\right)\right] \nonumber\\
&=&  -\frac{\iimg }{ \beta} \sum_{p\neq 0} \frac{1}{ ( \iimg w +\frac{ 2 \pi p}{\beta})}\bigg[e^{-\frac{1}{2}\beta k \left((\frac{2 \pi p}{\beta})^2 + w^2\right)} - 1 \bigg]
\label{cigarqm39}
\eea
After Poisson resummation with respect to $p$ (see equation \eqref{fourtranerfc1}) we obtain,
\begin{eqnarray}
\widehat W (\beta) =  \sum_n    e^{- \beta n w}  \bigg[ && - \frac{1 }{2}\sgn\left(\frac{n}{k}-w\right)  \bigg.\erfc\left(\sqrt{\frac{k\beta}{2}}\left|\frac{n}{k}-w\right|\right) 
+ \sgn(\beta n)\,  \Theta\left[w\left( \frac{n}{k}-w \right)\right]
\nonumber \\
&& 
 + \sgn(w \b) \Theta (n\b \, \sgn(w \b))\bigg]
\label{cigarqm40}
\end{eqnarray}
Now we can take the limit\footnote{The last two terms vanish in this limit which is easier to see before Poisson resummation.} $\beta\rightarrow 0 $ . 
\begin{eqnarray}\label{cigareta}
   \widehat W(0) =  \sum_n \frac{1}{2} \sgn\left(w -\frac{n}{k} \right) 
\end{eqnarray}
It is easy to check that $\widehat W(\infty)$ vanishes.  Using \eqref{etabeta}, we obtain
\be
    \eta (0) =2 ( \widehat W(0) -\widehat W(\infty))  =  \sum_n \sgn\left(w - \frac{n}{k}\right)
\label{etacigar001}
\ee
It is instructive to  compare this result with a  target space computation of the spectral asymmetry  of the  boundary operator $\cB$ on a boundary located  at $r = r_{c}$. Using the inverse vielbeins from \eqref{viel},
the  Dirac operator near the  boundary takes the form 
\begin{eqnarray}
\iimg \slashed{D}&=&	\gamma^r(\iimg  \partial_r -w\, K_r) +\gamma^\theta (\iimg \partial_\theta   -w\, K_\theta)
\nonumber\\
&=& i \gamma^{r}\bigg[ \partial_r -  \frac{1}{\tanh r}
\left(\begin{matrix}
	-1& 0\\
0	&1 \\
\end{matrix} \right) (\iimg \partial_\theta   -w\,k \tanh^2 r) \bigg]
\label{cigarqm13}
\end{eqnarray}
For large $r_{c}$, the boundary manifold is a circle $S^{1}$. Identifying $r$ with $u$ and comparing with \eqref{apsdirac} we find the    boundary operator
\begin{eqnarray}
\mathcal{B} 
&=&   - (\iimg \partial_\theta -w\,k ) \, .
\label{cigarqm14}
\end{eqnarray}
The $\eta$-invariant  of this operator can be computed readily.  Since $\theta$ direction is periodic,  the eigenfunctions are given by the set 
\be
\{  e^{-\iimg n \theta} | n \in \bZ \}
\ee
with eignevalues
\be
\{wk-n  | n \in \bZ \}
\ee
The radius of the cigar is $\sqrt{k}$. As long as $k$ is not an integer, the boundary operator $\cB$ has no zero modes. 
The $\eta$ invariant is then given by:
\begin{eqnarray} \label{etacigar}
    \eta = \sum_{n \in \bZ} \sgn ( w \, k-n) 
   =  \sum_{n \in \bZ} \sgn\left(w-\frac{n}{k} \right)
    \end{eqnarray}
where in the last step we have used the fact that $k$ is positive. This matches with the $\eta$ invariant computed from the   path integral \eqref{etacigar001}. 

In the infinite sum, one can absorb the integer part $\lfloor wk \rfloor $ of $wk$ into $n$,  and hence the $\eta$ invariant is expected to depend only on the  fractional part $\langle wk \rangle $  of $wk$  defined by
$$
	\langle wk \rangle = wk - \lfloor w k \rfloor 
\label{etaexample28}
$$
where $\lfloor wk \rfloor $ is the greatest integer  less that $wk$. 
The   regularized version of the $\eta$-invariant  is
\be
	\eta(s) = - \sum_{n=1}^\infty\frac{1}{(n- \langle wk \rangle)^s}+\sum_{n=0}^\infty\frac{1}{(n + \langle wk \rangle)^s} \, .
\label{etaexample30}
\ee
The $\eta$-invariant can now be expressed in terms  of the modified $\zeta$ function
\begin{eqnarray}
\zeta(s,q)=\sum_{n=0}^\infty \frac{1}{(n+q)^s}	
\qquad,\qquad
\zeta(0,q)=-q+\frac{1}{2}
\label{etaexample24}
\end{eqnarray}
to obtain
\be
\eta(0)  = -\zeta(0,1-\langle wk\rangle)+ \zeta(0,\langle wk \rangle)
= 1-2\langle wk \rangle 	
\ee
Note that for $k$ and $w$ both integers, the $\eta$-invariant vanishes.
As one varies $k$ the $\eta$-invariant changes and everytime $k$ crosses, an integer it jumps by $-2$. This is as expected from level-crossing because precisely when $k$ is an integer, the boundary operator has a zero eigenvalue. 
We also see that the case of a zero eigenvalue can be treated by slightly deforming the boundary operator but the answer depends on the direction in which one approaches zero.  This ambiguity is present also in the APS theorem and can be accounted for by assigning $\sgn(0)=1$ as in \eqref{etadef}. 
This behavior is plotted in  Figure \textbf{\ref{fig5}} with $\eta$ on the $y$-axis  and  $wk$ on the 
$x$-axis. 
\begin{figure}[h]
\begin{center}
\vskip4mm
\begin{tikzpicture}[ scale=1, color=blue ]
 
\draw[black] [ultra thick](-1,0) -- (6,0);

\draw[-] [thick](0,1) -- (1,-1);
\draw[-] [thick](1,1) -- (2,-1);
\draw[-] [thick](2,1) -- (3,-1);
\draw[-] [thick](3,1) -- (4,-1);
\draw[-] [thick](4,1) -- (5,-1);

\draw[dashed] [thick](1,1.5) -- (1,-1.5);
\draw[dashed] [thick](2,1.5) -- (2,-1.5);
\draw[dashed] [thick](3,1.5) -- (3,-1.5);
\draw[dashed] [thick](4,1.5) -- (4,-1.5);

\node[black] at (-1, -.5) {$wk $};

\end{tikzpicture}
\end{center}
\caption{Spectral asymmetry}
\label{fig5}
\end{figure}
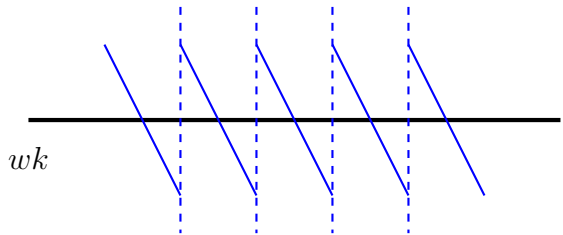

\section{Mock modularity and the $\eta$-invariant \label{Mock}}

In this section, we take $k$ to be a positive integer  and consider a worldsheet path integral for the superconformal field theory for an infinite cigar. The SCFT has a representation as a coset conformal field theory\footnote{For the noncompact $SL(2, \bR)$ WZW model, the parameter $k$ need not in general be an integer.} of   $SL(2, \bR)/U(1)$ WZW model at level $k$.   The elliptic genus for the $\mathbb{Z}_N$-orbifold of this theory was computed in \cite{Eguchi:2010cb, Ashok:2011cy} using the path integral for  the  coset theory
\be\label{eguchi}
\widehat\chi (\t, \bar\t| z)  = e^{\frac{2\pi}{\tau_2} \left( \widehat c |z|^2 - \frac{k +4}{k}z_2^2\right)}\frac{k}{N} \sum_{a,b \in \mathbb{Z}_N}\int_\mathbb{C} \frac{d^2u}{\tau_2} \, \left|\frac{\vartheta_1(\tau, -u + \left(1 + \frac{2}{k}\right)z)} {\vartheta_1(\tau, -u + \frac{2}{k}z)}\right|^2 e^{\frac{\pi k}{\tau_2}\left|u + \frac{a\tau +b}{N}\right|^2}
\ee
where  $\vartheta_1(\t,z)$ is  the odd Jacobi theta function
\begin{align}
\vartheta_1(\tau,z) = 2 \sin \pi z \, q^{1/8}  \prod_{n=1}^\infty
(1-q^n) (1- y q^n) (1- y^{-1}q^{n})\ .
\end{align}
We display the $\bar\t$ dependence explicitly to emphasize the nonholomorphicity.
The cigar elliptic genus was a computed in \cite{Troost:2010ud} using free field calculation and modular properties and was shown to be given by (in a notation slightly different from \cite{Troost:2010ud})
\be
\widehat\chi (\tau, \bar \t | z) =  -\frac{i\vartheta_{1}(\t,z) }{\eta(\t)^3} {\hat {\mathcal{A}}_{1, k}}\big(\t, \bar \t; \frac{z}{k}\big)
\label{resultccc}
\ee
where $\eta (\t)$ is the Dedekind eta function. The completion $\hat {\mathcal{A}}_{1,k}$ of  the Appel-Lerch sum ${\mathcal{A}}_{1,k}$
\begin{align}
{\mathcal{A}}_{1,k}(\t,z)= \sum_{t\in\mathbb{Z}}
\frac{q^{k t^2} y^{2kt}}{1-  y \, q^t}\ .
\label{ES}
\end{align}
is given by

\be
\widehat{\mathcal{{A}}}_{1,k}\left(\t, \bar \t; z\right) = \mathcal{{A}}_{1,k}\left(\t, z \right) + \sum_{\mathbb{Z}/2m\mathbb{Z}}   g^*_\ell(\tau , \bar \t) \, \vartheta_{k,l} \left(\tau, z\right)
\ee
with
\be \label{gcompletion}
 g_{\ell}^* (\tau, \bar{\tau})\, =  - \frac{1}{2}\sum_{ {r = \ell +2k\mathbb{Z}}}  \sgn\left(r\right) \erfc\left(|r|\sqrt{\frac{ \pi \tau_2}{ k}}\right) q^{-r^2/4 k}  
\ee
It was shown in \cite{Eguchi:2010cb} that  \eqref{eguchi} agrees with \eqref{resultccc} for $N = 1$.
 The same result was obtained in \cite{Murthy:2013mya,Ashok:2013pya} using localization of the generalized linear sigma model that flows to the cigar geometry. The cigar elliptic genus was re-derived in \cite{Giveon:2015raa,Troost:2017fpk} using canonical methods.  
 
 Our aim is to re-re derive the elliptic genus of an infinite cigar using localization of the  path integral directly in the nonlinear sigma model connecting the computation to the earlier discussion of the  $\eta$-invariant for the boundary of the finite cigar. For this purpose,  it is convenient to re-express \cite{Dabholkar:2012nd}   the Appell-Lerch sum in the range $|q|<|y|<1$ as 
 \be
{\mathcal{A}}_{1,k}(\t,z) = \bigg(- \sum _{s\geq0} {\sum_{l\geq0}}^* + \sum _{s<0} {\sum_{l<0}}^*\bigg)  q^{ks^2+ sl} y^{2ks+l}
\ee
where the asterisk in the sum indicates that the term $\ell = 0$ is counted with multiplicity 1/2. This is the form in which we will encounter it in \S \ref{CigarEll}.

\subsection{Mock Jacobi forms \label{MockJ}}

In \S \ref{Elliptic} we defined the elliptic genus of a compact SCFT. 
One can generalize the definition to a noncompact SCFT by generalizing the trace  to include the scattering states as for the noncompact Witten index
\begin{equation}\label{noncompactchi}
\widehat\chi (\t, \bar\t| z) =\TrS (-1)^{\tilde J + J} {e}^{2 \pi i \tau L_{0}} e^{-2 \pi i \bar{\tau} \tilde L_{0}} e^{2 \pi i z J} \, .
\end{equation}
One can then write a path integral representation of the elliptic genus which can be a starting point for using localization methods. 
Even in the noncompact case, the path integral of the SCFT with a covariant regulator is  manifestly diffeomorphism and Weyl invariant. Hence it  is expected to be modular. Moreover, the spectral flow symmetry for the current $J$ guarantees that it will be elliptic. 
Consequently, by the arguments outlined in \S\eqref{Elliptic},  we expect the elliptic genus to transform as a Jacobi form of weight $0$ and index $m$ as in the compact case. However, unlike in the compact case, it need not be independent of $\bar \tau$. The elliptic genus for the cigar turns out to be  a \textit{completion of a mixed mock Jacobi form} which is nonholomorphic but transforms like a Jacobi form. 

We  recall a few definitions about (mixed) mock Jacobi forms and their completions \footnote{Our definitions  are a slight variant of the definitions in \cite{Dabholkar:2012nd}, better suited to the problem at hand.}.  A completion of a mixed mock Jacobi form admits the following theta expansion:
\be 
\label{noncompactphi}
    \widehat{\phi} (\tau, \bar \tau| z) = f(\tau, z) \sum_{l \, \in  \mathbb{Z}/2m\mathbb{Z}}\widehat h_l (\tau, \bar \tau) \, \vartheta_{k,l}(\tau|z)
\ee
where $f(\t, z)$ is a jacobi form of weight $u$ and index $\alpha$. The theta expansion can once again be seen by bosonizing  the current $J$ as in the compact case. The theta coefficients are the  \textit{ completion of a  vector valued mock modular form} of weight $(w-u -\half)$. The nonholomorphic sum $\widehat h_{\ell}(\t, \bar \t) := h_{\ell}(\t) \+  g_{\ell}^*(\t, \bar \t)$ transforms like a vector-valued modular form where $g_{\ell}^*(\t, \bar \t)$  is a solution of the following differential equation.
\be\label{starinv} (4\pi\t_2)^v\,\frac{\pa g_{\ell}^*(\t, \bar \t)}{\pa \bar{\tau}} \= -2\pi i\;\widebar{g_{\ell}(\t)} \, . \ee
where $v = w - u-\half$. The holomorphic modular form $g_{l}(\t)$ of  weight $2 -v$. Using the completion of $\hat h_\ell$, \eqref{noncompactphi} can be written as
\begin{eqnarray}\label{jacobicompletion}
    \widehat{\phi} (\tau, \bar \tau| z) = \phi(\tau,   z) + f(\tau, z) \sum_{\ell\in  \mathbb{Z}/2m\mathbb{Z}}   g^*_\ell(\tau , \bar \t) \, \vartheta_{m,\ell}(\tau |z) 
 \end{eqnarray}
where
\be 
\phi (\tau,  z) = f(\tau, z) \sum_{l \, \in  \mathbb{Z}/2m\mathbb{Z}} h_l (\tau) \,\vartheta_{k,l}(\tau|z)
\ee
Given that $g_{\ell} (\t)$ has the Fourier expansion $g(\t)=\sum_{n\ge0}\;b_{\ell, n}\,q^n$, we fix the choice of $g_{\ell}^*$ by setting 
\be\label{defstar} 
  g_{\ell}^*(\t, \bar \t)  \=  \bar b_0\,\frac{(4\pi\t_2)^{-v+1}}{v-1} \+ \sum_{n>0} \;n^{v-1}\,\bar b_{\ell, n} \;\G(1-v,4\pi n\t_2)\;q^{-n}\, , 
\ee
where $\t_2 = \rm{Im}(\t)$ and $\G(1-v,x)$  denotes the incomplete gamma function defined as in \eqref{incomplete}, and where 
the first term must be replaced by $ -\bar b_0\,\log(4\pi\t_2)\,$ if $v=1$. 

Note that the series in~\eqref{defstar} converges despite 
the exponentially large factor $q^{-n}$ because $\G(1-v,x)= O(x^{-v}e^{-x})\,$. If we assume either that $v>1$ or that $b_{0}=0$,
then we can define $g_{\ell}^*$ alternatively by the integral  
\be\label{Lkintrep} g_{\ell}^*(\t, \bar\t) = \biggl(\frac i{2\pi}\Bigr)^{v-1} \int_{-\bar{\t}}^{\infty} (z+ \t)^{-v} \ \widebar{g(-\bar{z})}\; dz \;. \ee
(The integral is independent of the path chosen because the integrand is holomorphic in $z$.) 
Since $\phi(\t,z)$ is holomorphic, \eqref{starinv} implies that the completion of $h$ is related to its shadow by
\be\label{ddtbarh}   (4\pi\t_2)^v\,\,\frac{\pa \widehat \phi(\t, \bar \t | z)}{\pa \bar{\tau}} \= -2\pi i\; f(\tau, z) \sum_{\ell\in  \mathbb{Z}/2m\mathbb{Z}}   \widebar{g_{l}(\tau)} \, \vartheta_{m,\ell}(\tau |z) \;.  
\ee
We will apply these definitions in the next section to the special case of our interest.

\subsection{Elliptic genus of an infinite cigar \label{CigarEll}}

The computation of $\eta$-invariant for a finite cigar can be used to compute the full elliptic genus for a $N=(2,2)$ SCFT  on an infinite cigar \footnote{The $SL(2, \bR)/U(1)$ coset has  is a non-trivial background `spacetime' dilaton  
 \be
 \Phi_{d}(r) = \Phi_{d0} - \log \cosh r
 \ee
which ensures that  the theory is conformal even though the target space is not Ricci flat.  Since the dilaton  couples to the worldsheet curvature, it plays no role if the worldsheet is a torus as in our case.}. 
Notice that for the cigar case, the $R$-symmetry generator $J$ appearing in the definition of elliptic genus \eqref{noncompactchi} commutes with the right moving supercharge. Hence for the right movers the computation reduces to computing the non-compact Witten index  computed in  \S\ref{Cigar}. The full elliptic genus is given by
\begin{eqnarray}\label{etagenus}
\widehat\chi (\t, \bar \t | z) 
&=&
\TrH[] (-1)^{\tilde J} {e}^{-2 \pi \tau_2(L_{0}+ \tilde L_0)} e^{2 \pi i {\tau}_1 (L_{0} - \tilde L_0)} e^{2 \pi i z J} \nonumber\\
&=& \widehat W(2\pi \tau_2) \cdot  \mathcal{Z}_{\textrm{oscill}}\,\,  e^{2 \pi \iimg \tau_1 m w}\, e^{2 \pi i z J}
\end{eqnarray}
where $\widehat W(2\pi \tau_2)$ is the Witten index with $\beta = 2\pi \tau_2$, $\mathcal{Z}_{\textrm{oscill}}$ is the contribution coming from left-moving oscillators and $n$, $w$ are KK momenta and winding respectively along the cigar $\theta$ direction. The contribution coming from the oscillators is given by
\be 
\mathcal{Z}_{\textrm{oscill}} = \prod_{n = 1}^{\infty} \left[\frac{(1 - q^n y)(1 - q^n y^{-1})}{(1 - q^n)^2}\right] = (2 \sin \pi z)^{-1} \frac{\theta_1(\t,z)}{\eta(\t)^3}
\ee 
  We get a contribution of $(2 \iimg \sin \pi z)$ from the zero modes of left moving fermions since they are charged under $U(1)_R$. Now we can substitute the contribution. Using equation \eqref{cigarqm40} and the contribution from left movers, we conclude that the elliptic genus for the cigar  is given by 
\begin{eqnarray}\label{Cigaregfull}
\widehat\chi (\t, \bar \t | z) 
&=& -i \frac{\vartheta_1(\t, z)}{\eta^3(\tau)} \sum _w \sum_n \bigg[\frac{1}{2}\sgn\left(\frac{n}{k}-w\right) \erfc\left(\sqrt{k \pi \tau_2}\left|w - \frac{n}{ k}\right|\right)\bigg. \nonumber\\ 
 &&- \bigg.  \sgn(\beta n)\,  \Theta\left[w\left( \frac{n}{k}-w \right)\right]
 \bigg]  
 q^{-(n-wk)^2/4 k} q^{(n+wk)^{2}/4 k} y^{J_L}
\end{eqnarray}
To obtain the above expression we have dropped the last term in \eqref{cigarqm40} using the following reasoning. At the tip of the cigar an infinite number of winding modes become massless leading to a divergence for this term.  This is a consequence of the fact that winding number is strictly not a conserved quantum number at $r=0$ as we have assumed.   We can deal with it by regularizing the Witten index $\widehat W (2\pi\tau_2)$ near $r=0$ by putting an $\epsilon$ cutoff in the $r_0$ integral in \eqref{cigarqm36} and then taking $\epsilon \rightarrow 0$ in the end.  With this regularization, the contribution from the last term in \eqref{cigarqm40} vanishes and we get (somewhat surprisingly)  the correct answer by this slightly heuristic procedure. In any case, this affects only the holomorphic piece and not  the holomorphic anomaly which is our main interest.  Since the holomorphic anomaly is determined by  the scattering states,  winding-number in the asymptotic region  is a good quantum number for our purposes. As a result the holomorphic anomaly is not affected by this regularization.

Note that on the cigar,  the $R$-current is given by
\be
J = i\sqrt{\frac{1}{k}}\partial \theta - i \psi_r\psi_\theta
\ee
and as a consequence not only the fermions but bosons are also charged under $R$-symmetry. With this normalization\footnote{ We use $\alpha' = 1$  so that asymptotic radius $R$ of the cigar is $\sqrt{k}$ while \cite{Troost:2010ud} uses $\alpha' = \sqrt{2}$.} \cite{Troost:2010ud}, the left-moving fermions have charge $-1$ and the bosons have charge $1/k$. In terms of the left moving momenta, the $R$-current is given by $J = \sqrt{1/ k} \, p_L$. The left and right moving momentas are given by:  
\be
p_{L}=   \left(\frac{n}{R} + w R\right) \, \qquad,\qquad p_{R} =  \left(\frac{n}{R} - w R\right)
\ee
The expression for the elliptic genus  is non-holomorphic but it is modular if  $k$  is an integer. More precisely, it transforms as a \textit{completion} of a mock Jacobi form of weight $0$ and index $m$.  

The holomorphic piece is given by:
\be
\widehat\chi_h (\t, \bar \t | z) 
= i \frac{\vartheta_1(\tau,z)}{\eta^3(\tau)}\bigg[\sum _{w\geq0} \sum_{n - wk\geq0} - \sum _{w<0} \sum_{n - wk<0}\bigg] q^{nw} y^{\frac{n + wk}{k}}
\ee
We can replace the sum in \eqref{Cigaregfull} over by $n$ and $w$ by the sum over $s$ and $s'$ by noting that 
\be \ell = (n + wk) = (n -wk) \; \mod \; 2k \,
.  
\ee 
Or equivalently 
\be n-wk = \ell + 2ks \, \qquad,\qquad n + wk = \ell+ 2ks' \, .  \ee 
Hence the holomorphic piece can be written as
\be\label{cigarholomorphic}
\widehat\chi_h (\t, z) = i \frac{\vartheta_1(\t, z)}{\eta^3(\t)}\bigg[\sum _{w\geq 0} \sum_{\ell \geq 0} - \sum _{w<0} \sum_{\ell<0}\bigg]q^{w^2k + w \ell } y^{\frac{\ell+2kw}{k}} = -i \frac{\vartheta_1(\t, z)}{\eta^3(\t)} \mathcal{A}_{1,k}\left(\t,\frac{z}{k}\right)
\ee
which matches with the result obtained in \cite{Troost:2010ud}.  

To find the shadow, let us focus only on the non-holomorphic piece $\widehat\chi_{nh} (\t, \bar \t | z) $ which equals
\begin{eqnarray}\label{CigarEGtheta} 
&& - i \frac{\vartheta_1(\t, z)}{\eta^3}\sum_{l \, \in  \mathbb{Z}/2k\mathbb{Z}} \sum_{s, s'} \frac{1}{2}  \sgn\left(\ell + 2ks \right) \erfc\left(|\ell + 2ks|\sqrt{\frac{ \pi \tau_2}{ k}}\right) q^{-(\ell + 2ks)^2/4 k} q^{(\ell+ 2ks' )^{2}/4 k} y^{\frac{\ell+ 2ks'}{k} } \nonumber\\
 &=& -i \frac{\vartheta_1(\t, z)}{\eta(\t)^3}\sum_{l \, \in  \mathbb{Z}/2k\mathbb{Z}} \,  \sum_{ {r = \ell +2k\mathbb{Z}}}\frac{1}{2}  \sgn\left(r\right) \erfc\left(|r|\sqrt{\frac{ \pi \tau_2}{ k}}\right) q^{-r^2/4 k} \vartheta_{k,l} \left(\tau,\frac{z}{k}\right)
\end{eqnarray}
Combining the holomorphic \eqref{cigarholomorphic} and non-holomorphic  \eqref{CigarEGtheta} contribution, the elliptic genus of cigar is given by
\be
\widehat\chi(\t, z) =- i \frac{\vartheta_1(\t, z)}{\eta^3(\t)} \widehat{\mathcal{{A}}}_{1,k}\left(\t,\frac{z}{k}\right)
\ee
Comparing \eqref{CigarEGtheta} and \eqref{jacobicompletion} we conclude that
our elliptic genus is a mixed mock Jacobi form with
\be
f(\t, z) = -i \frac{\vartheta_{1}(\t, z)}{\eta^{3}(\tau)} \, , \qquad \qquad g_{\ell}^* (\tau, \bar{\tau})\, = \sum_{r = \ell+ 2k\mathbb{Z}}  \frac{1}{2} \sgn(r)\,  \erfc\left(|r|\sqrt{\frac{ \pi \tau_2}{ k}}\right) q^{-r^2/4 k} \, .
\ee 
and with  $w=0$ $u = -1$ and hence $v = 1/2$.    
The total  index is $m = \frac{1}{2} + \frac{1}{k}$ which matches\footnote{Note that $\widehat{\cA}_{1,k}(\t,z)$ is a Jacobi form with index $k$ but $\widehat{\mathcal{{A}}}_{1,k}\left(\t,\frac{z}{k}\right)$ has index $\frac{1}{k}$ because of the rescaling of $z$. The theta function $\vartheta_1(\t, z)$ transforms with index $\half$.} with the expected index $m = c/6$ where $c$ is the central charge of the coset.
 We can now compute the holomorphic anomaly using \eqref{ddtbarh} with $v=\half$
\be
(4\pi\t_2)^{1/2}\,\,\frac{\pa \widehat \chi (\t,\bar \t|z)}{\pa \bar{\tau}}
= -\sqrt{\frac{\pi}{2 k}} \frac{\vartheta_1(\t, z)}{\eta^3(\tau)}  \sum_{\ell\in  \mathbb{Z}/2k\mathbb{Z}}\overline {\vartheta^{(1)}_{k,l}(\tau)}\vartheta_{k,l} \left(\tau,\frac{z}{k}\right)
\label{cigareg105}
\ee
where $\vartheta^{(1)}_{k,l}$ is the unary theta function which is defined as
\begin{eqnarray}\label{unary1}
    \vartheta^{(1)}_{k,l}(\tau) = \frac{1}{2 \pi \iimg}\frac{d}{dz}\vartheta_{k,l} (\tau,z)\bigg|_{z = 0} = \sum_{r \equiv \ell \, (\mod \; 2m)}r \, q^{r ^{2}/4 k}
\end{eqnarray}
Comparing \eqref{ddtbarh} and \eqref{cigareg105} we find that shadow vector is 
\be
g_{l}(\tau) = -\frac{1}{\sqrt{8\pi k}}\vartheta^{(1)}_{k,l}(\tau) 
\ee
The shadow vector $\{g_{l}(\t)\} $ is an unary theta series as in \eqref{unary1}. In this case with $v=\half$,  the incomplete gamma function
in~\eqref{defstar} can be expressed in terms of the complementary error function $\,\erfc(x)$ using \eqref{icerror}. 

In an interesting recent paper \cite{Gaiotto:2019gef} the holomorphic anomaly was related to a one-point function on the torus, which by our analysis should be determined by the scattering states.

\subsection{ The $\eta$-invariant  and quantum modular forms \label{Quantum}}

We consider the radial limit ($\tau_2\rightarrow 0^{+} $) of the non-holomophic part $g^\star_\ell(\tau,\bar \tau)$ in \eqref{gcompletion} to obtain
 \be
 \label{etagstar}
\lim_{\tau_2 \rightarrow 0} g_{\ell}^* (\tau, \bar{\tau})\, = - \frac{1}{2} f_{\ell} (\t_{1}) \, , 
\qquad\qquad
\tau_1\in \mathbb{Q}
\ee
where 
\be
f_{\ell} (\t_{1})  = \sum_{ {r = \ell + 2k\mathbb{Z}}}  \sgn\left(r\right)  e^{- 2 \pi \iimg \tau_1 r^2/4 k} \, .
\ee
The characteristic $\sgn(r)$ that appears in the $\eta$-invariant appears here too but now in a sum  weighted with a phase.  This infinite sum is not   convergent  but can be regularized to get a finite answer  at rational  points on the real line. Thus, the function is defined only over $\bQ$ and not over $\bR$.  However, the difference between the function and its modular transform has  `nice' properties over $\bR$. 
More precisely, the vector $f(\t_{1}) $  transforms as
\be
f(\t_1) - (c \t_{1} + d)^w f\left(\frac{a \t_{1} + b}{c \t_{1} + d}\right) = h_\gamma (\t_1)
\ee
where 
$ \gamma = \begin{pmatrix}
	a     & b \\
	c      & d
\end{pmatrix} \in SL(2,\mathbb {Z})$ and the `period integral'  $h_{\g}(\t_{1})$ has some property of continuity  and analyticity for every element $\gamma$ and $\t_{1}\in \bR$.  
In  the terminology of  \cite{zagier2010quantum}, 
$\{f_{\ell}(\t_{1})\}$ transform  as  (vector-valued) `quantum modular forms'  of weight $\half$ and are naturally regarded as the theta-coefficients of a `quantum  Jacobi form' \cite{Bringmann_2016}.

We obtained the quantum modular form as a limit of the completion of  a mock modular form defined on the upper half $\tau$ plane $\bH^{+}\,  (\t_{2}>0)$  but it can equally well be obtained as a limit of a \textit{false} theta function  defined on the lower half plane $\bH^{-}\,  (\t_{2} < 0)$. Consider the false theta function \cite{MR2474043,Creutzig:2013zza}
} defined by
\begin{eqnarray}
    F_{a, b}(\tilde q) = \sum_{n\in \mathbb{Z}}\sgn(n)\, \tilde q^{a(n + \frac{b}{2a})^{2}} \quad  \quad (\tilde q:= e^{-2\pi i \tau}) \, .
\end{eqnarray}
With $r = \ell + 2 k  n$,  one can rewrite the $f_{\ell}(\t_{1})$ as
\be
f_{\ell} (\t_{1})  = \sum_{n\inn \bZ}  \sgn\left(\ell + 2 k  n\right)  e^{- 2 \pi \iimg \tau_1( \ell + 2 k  n)^2/4 k}
\ee
which can be viewed as a $\t_{2} \rightarrow 0^{-}$ limit of the false theta function for $a =k$ and $b =l$ using the fact that $\sgn\left(\ell + 2 k  n\right) = \sgn(n)$ for positive $k$ and $0 \leq l < 2k$. In this limit, the functions $f_{\ell}(\t_{1})$ have appeared in the computation of topological invariants of Seifert manifolds with three singular fibers \cite{Cheng:2018vpl}. See \cite{Cheng:2018vpl} for a  discussion of quantum modular forms  and the relation to mock and false theta functions in the context of Chern-Simons theory and WRT invariants \cite{Witten:1988hf, Reshetikhin:1991tc, lawrence1999modular, Gukov:2017kmk, bringmann2018quantum}.
Further discussion of quantum modular forms can be found in   \cite{Zwegers:2008zna, bringmann2018quantum}.  It would be interesting the explore further these connections with the nonholomorphic  elliptic genus.

Motivated by our expression  for the $\eta$-invariant \eqref{etabeta}, one can consider
\be 
\eta(\t_{1}|z) = 2\bigg[\widehat \chi(\tau, \bar \t |z)\big|_{\t_{2}\rightarrow 0} - \widehat \chi(\tau, \bar \t |z)\big|_{\t_{2}\rightarrow \infty}\bigg]
\label{etagenus1} \, .
\ee 
Going back to the definition of the elliptic genus in terms of the trace, we can interpret $\eta(\t_{1}|z)$  as a particular  `character-valued' $\eta$-invariant of the elliptic genus
\begin{eqnarray}
\eta(\tau_1 |z) &=& \mathcal{Z}_{\textrm{oscill}} \sum_w  2 \left(\widehat W(0) - \widehat W(\infty)\right) e^{2 \pi i \tau_1 (L_0 - \bar L_0)} y^{J} \, . 
\end{eqnarray}
Evaluating the limits, we find
\begin{eqnarray} 
\lim_{\tau_2 \rightarrow 0} \widehat \chi(\tau, \bar \t |z) &=&- i \frac{\theta_1(\tau_1, z)}{\eta(\tau_1)^3} \sum_w \sum_n\left(-\sgn(n) + \frac{1}{2} \sgn \left(\frac{n}{k} - w \right)\right) e^{2 \pi i \tau_1 nw} y^{\frac{n+wk}{k}} \label{limitchi} \\
\lim_{\tau_2 \rightarrow \infty} \widehat \chi(\tau, \bar \t |z) &=& 0
\label{limits}
\end{eqnarray}
Substituting into \eqref{etagenus1}, the `character-valued $\eta$-genus' takes the form
\be\label{etagenus}
\eta(\t_{1}|z) 
= -2 i \frac{\vartheta_1(\t_1, z)}{\eta(\t_1)^3}\left[ \sum_{l \, \in  \mathbb{Z}/2k\mathbb{Z}} \, f_{\ell} (\t_{1}) \vartheta_{k,l}\left(\tau_1,\frac{z}{k}\right) + \cA_{1,k}\left(\tau_1,\frac{z}{k}\right)  \right] \, .
\ee
The second term must also transform `nicely' with the same  period integrals for its theta coefficients up to a sign, consistent with the fact that $\widehat \chi(\tau, \bar \t |z)$  transforms as a Jacobi form. 
One peculiarity of the cigar conformal field theory is that the $R$-symmetry acts not only the fermions but also on the bosons by shifting  the theta coordinate of the cigar. This is different from what happens for the compact elliptic genus. Consequently,  the precise interpretation of this $\eta$-invariant from the perspective of a tower of Dirac operators is not completely clear to us. 

If we think of the elliptic genus as a right-moving Witten index, another perhaps more    natural limit is to consider the $\bar \tau \rightarrow 0$  to obtain
 \begin{eqnarray}
\lim_{\bar \tau \rightarrow 0} g_{\ell}^* (\tau, \bar{\tau})\,&=&   - \frac{1}{2} \sum_{r = \ell + 2mk}  \sgn\left(r\right) \erfc\left(|r|\sqrt{\frac{ \pi \tau}{2 \iimg k}}\right) q^{-r^2/4 k} \, .
\end{eqnarray}
The $\bar \tau \rightarrow -i \infty$ limit gives zero.  It is not clear to us how to interpret this expression.

\subsection*{Acknowledgments}

We thank Gurbir Singh Arora, Andr\'e Benevides, Nima Doroud,  Jeffrey Harvey, Pavel Putrov, T Ramadas, Jan Troost, and especially Francesca Ferrari for useful discussions. 
\appendix
\section{Appendix}
\subsection{Conventions and $(1, 1)$ superspace \label{Spinor}}

We use  indices $\{ \a, \b, \ldots \}$ with values in $ \{ 0, 1 \}$ to label the components of a worldsheet vector  and   $\{ A, B, \ldots \}$ with values in $\{ +, -\}$  to label the components of a worldsheet spinor.  On field space, we use $1 \leq i , j, \ldots \leq 2n$ as the coordinate indices  $1 \leq a, b, \ldots \leq 2n$ as the tangent space indices. We use the worldsheet metric to be $\eta _{\alpha \beta} = \textrm{diag}(-,+)$. 

A convenient basis for the two dimensional Dirac matrices is
 
\begin{eqnarray}
\rho^0 
=\left(
\begin{matrix}
0 & -\iimg  \\
\iimg & 0 
\end{matrix}  
\right)
\qquad,\qquad
\rho^1 =
\left(
\begin{matrix}
0  & \iimg  \\ 
\iimg  & 0
\end{matrix}
\right)
\label{twodimsusy3}
\end{eqnarray}
which satisfy $\{\rho^\alpha, \rho^\beta\} = -2 \eta^{\alpha \beta}$. The worldsheet chirality $\bar \rho$ and charge conjugation matrix $C$ are 
\begin{eqnarray}
\bar \rho = -\rho_0\rho_1 =\left(
\begin{matrix}
-1 & 0 \\
0 & 1
\end{matrix}
\right) \, \qquad \mathcal{C}_{AB}=	\rho^0_{AB} 
\label{twodimsusy4}
\end{eqnarray}

A Majorana spinor we mean a two-component real spinor 
The usual definition of Majorana spinor is that
$ \bar \psi= \psi^\dagger \mathcal{C}$. In this case, $\mathcal{C}=\rho^0$ and hence you get the above condition.  
\begin{eqnarray}
\psi =    \begin{pmatrix}
\psi_-\\
\psi_+
\end{pmatrix} \, \qquad \psi=\psi^*	
\end{eqnarray}
We use the superspace $s\Sigma$ with real superspace coordinates $\{ \s^{\a} , \theta_{A}\}$ to write down the supersymmertic lagrangian. We use the following convention for superspace derivatives and integrals
\be
\frac{\partial}{\partial \theta^A}\theta^B = \delta_A^B , \qquad \frac{\partial}{\partial \bar\theta^A}\bar \theta^B = \delta_A^B \qquad \textrm{and} \qquad \int d\theta d \bar{\theta} \, \, \bar{\theta} \theta = 1.
\ee
In superspace, the supercharge is given by
\be
\scharge_\ssa=
\frac{\partial}{\partial \bar \theta^\ssa }	
+\iimg (\rho^\wsa \theta)_\ssa \partial_\wsa 
\ee
which satisfy the  $\mathcal{N}=(1,1)$ supersymmetry algebra 
\begin{eqnarray}
\{\scharge_\ssa,\scharge_\ssb\}=	2\iimg (\rho^\wsa \partial_\wsa )_{\ssa \ssb}
\label{twodimsusy22}
\end{eqnarray}
To write actions invariant under the supersymmetry, one needs a supercovariant derivative. Supercovariant derivative is invariant under supersymmetry and it is defined by
\begin{eqnarray}
\spartial_\ssa=
\frac{\partial}{\partial \bar \theta^\ssa }	
-\iimg (\rho^\wsa \theta)_\ssa \partial_\wsa 
\label{twodimsusy21}
\end{eqnarray}
It satisfies  the following anticommutations
\begin{eqnarray}
\{\spartial_\ssa,\spartial_\ssb\}=	-2\iimg (\rho^\wsa \partial_\wsa )_{\ssa \ssb}
\label{twodimsusy23}
\end{eqnarray}
Since  $\bar \rho_{\ssa \ssb }$ is symmetric, it allows for   a central extension of the supersymmetry
\begin{eqnarray}
\{\widehat {\scharge}_A,\widehat\scharge_B\}=	2\iimg (\rho^\wsa \partial_\wsa )_{\ssa \ssb}
+ 2\iimg (\rho)_{\ssa \ssb}
\label{twodimsusy28}
\end{eqnarray}
In presence of central charge the super-charge and the covariant derivative are modified as 
\begin{eqnarray}
{\widehat \scharge}_\ssa =\scharge_\ssa + \iimg (\bar \rho\, \theta)_\ssa Z
\qquad,\qquad
\widehat \spartial_\ssa =\spartial_\ssa -\iimg (\bar \rho\, \theta)_\ssa Z
\label{twodimsusy29}
\end{eqnarray}
The presence of the killing vector in the target space allows a central term in the supersymmetry algebra. The central charge is related to the killing vector as $K_i = Z x_i$. The action \eqref{action2} modifies to
\begin{equation}
I \longrightarrow I + g_{ij}K^iK^j + \iimg \bar{\psi^i}D_j K_i \bar \rho \psi_j                                                                                    
\label{susysigma36} 
\end{equation}
where $D_j$ is a covariant derivative on $\cM$ \cite{AlvarezGaume:1983ab}. For an off-shell formulation see \cite{Gates:1983py}.

\subsection{Error function and incomplete Gamma function}

The error function and the complementary error functions are defined by
\be
\erf(z) := \frac{2}{\sqrt{\pi}}\int_{0}^{z} dy \, e^{-y^{2}} \,  , \qquad \erfc(z) := \frac{2}{\sqrt{\pi}}\int_{z}^{\infty} dy\,  e^{-y^{2}} \, .
\ee
They satisfy the following relation
\be
\erfc(z)  = 1- \erf(z)
\ee
Note that $\erf(z)$ is an odd function because the integrand is an even function
\be
\erf(-z) = -\erf(z) \, .
\ee
For the purpose of this paper it is convenient to use the expression 
\begin{eqnarray}\label{error5}
\erf(z) &=&\sgn(z)\,
\erf(|z|) \, , \nonumber \\
\erfc (z) &=&	1-\sgn(z)\, \erf(|z|)
\end{eqnarray}
for $ z\in \mathbb{R}$ to make contact with the $\eta$-invariant.

The upper incomplete Gamma function encountered in \S \ref{MockJ} is defined  by
\be\label{incomplete}
\G(s, x) = \int_x^{\infty} t^{s-1}\,e^{-t}\,dt \, , \qquad x \geq 0 \, .
\ee
A special case that we encounter is
\be\label{icerror}
\G (\half, x ) =  \sqrt{\pi}\, \erfc{(\sqrt{x})} \, .
\ee
One of the integrals (involving error function) which is useful in our computation is the following
\begin{eqnarray}
f(m)&=&  -\frac{\iimg }{ \beta}\int dp \frac{1}{ (  {\iimg} w +\frac{ 2\pi p }{\beta})}\bigg[\exp^{-\frac{1}{2}\beta k \left(\frac{(2\pi p)^2 }{\beta^2} \right) - 2\pi \iimg n \cdot p} \bigg]
\nonumber\\
&=&
-\frac{1}{2}  \sgn\left(\frac{n}{k}-w\right)\, \erfc\left( \sqrt{\frac{\beta k}{2}}\left|w- \frac{n}{k}\right| \right) e^{-\beta n w +\frac{\beta k}{2} w^2}
 +\sgn(\beta n)\,  \Theta\left[w\left( \frac{n}{k}-w \right)\right]\, e^{ -\beta n w   }
\nonumber\\ 
\label{fourtranerfc1}
\end{eqnarray}

\subsection{Scattering theory \label{Spectral}}

We review how the density of states can be related to the phase shifts in scattering theory. 
Consider the scattering problem for the Hamiltonian (See for example \cite{Sakurai:1167961})  
\begin{eqnarray}
	H=H_0+V e^{-\e |t|}
\label{spectraltheory11}
\end{eqnarray}   
where we have added an adiabatic switching factor for the interaction $V$ so that in the far past and and in the far future one obtains the free Hamiltonian $H_{0}$. 	
The time evolution operator in the  Dirac picture  is given by 
\begin{eqnarray}
	U_D(t, t')=e^{\iimg H_0 t  }\, U(t, t')\, e^{-\iimg H_0 t' }
\label{spectraltheory12}		
\end{eqnarray}
where $U(t, t')$ is the time evolution operator of the Heisenberg picture. 
The Dirac evolution operator  satisfies the Schro\"odinger equation
\begin{eqnarray}	
	\iimg \frac{d}{dt}U_D(t, t') =V_D(t)\, U_D(t, t') \, ,  \qquad with \qquad
		V_D(t)  =e^{\iimg H_0t  }\, V e^{-\e |t|} \, e^{-\iimg H_0 t  }
\label{spectraltheory13}		
\end{eqnarray}
with the initial condition $U_{D}(t, t) = \bf{1}$.  The solution is given by
\be\label{Usolution}
U_D(t, t') = \textbf{1} -i \int_{t'}^{t}dt'' V(t'') U_D(t'', t')
\ee
 We can now define the `M\"{o}ller operators'
\begin{eqnarray}
	U_{\pm }=U_D(0,\pm \infty) 
\label{spectraltheory14}		
\end{eqnarray}
Consider and energy eigenstate $|\phi_E\rangle$ of the free Hamiltonian $H_{0}$. Using the M\"{o}ller operators  one can obtain the  eigenstate of the full hamiltonian:
\begin{eqnarray}
	|\psi_E^\pm \rangle=U_{\pm }|\phi_E\rangle
\label{spectraltheory15}		
\end{eqnarray}
where $|\psi_E^- \rangle$ are  the \textit{in}-states that resemble the free eigenstates in the far past and $|\psi_E^- \rangle$ are the  \textit{out}-states that resemble the free eigenstates in the far future.  
Solving \eqref{Usolution} recursively gives the  Dyson series expansion
\be
U_{\pm} |\phi_E \rangle = |\phi_E \rangle  + \frac{V}{E-H_{0}\mp i\e}|\phi_E \rangle +  \left( \frac{V}{E-H_{0}\mp i\e} \right)^{2}|\phi_E \rangle + \ldots
\ee 
This geometric series can be easily summed to obtain
\begin{eqnarray}
	U_{\pm } |\phi_E \rangle = \frac{E-H_0\mp \iimg \epsilon }{E-H\mp  \iimg \epsilon }|\phi_E \rangle
\label{spectraltheory16}		
\end{eqnarray}
It follows that $|\psi_E^\pm \rangle$  satisfy the Lippman-Schwinger equations
\be
|\psi_E^\pm \rangle= |\phi_E\rangle + 
\frac{V}{E - H_{0} \mp i\e} |\phi_E\rangle \, .
\ee
The  $S$-matrix in the interaction picture is just the time evolution operator 
$U_{D}(\infty, -\infty)$ which can be expressed  in terms of the M\"{o}ller operators as 
\begin{eqnarray}
	S= U_+^\dagger\,  U_- \, .
\label{spectraltheory17}		
\end{eqnarray}
The derivative of the  $S$-matrix is given by
\begin{eqnarray}
	\frac{d\, \ln\,  S}{dE}= S^{-1}\frac{d  S}{dE}=S^\dagger \frac{d  S}{dE}
\label{spectraltheory18}		
\end{eqnarray}
and using the above formula it is possible show that
\begin{eqnarray}
	\frac{d\, \ln\,  S}{dE}
	=2\pi \iimg \rho(E)=2\pi \iimg\,  \Big[ \delta(E-H)- \delta(E-H_0)\Big]
\label{spectraltheory19}		
\end{eqnarray}
The density of states is then given by the so called `Krein-Friedel-Lloyd' formula:
\begin{eqnarray}
	\rho(E)=\Tr \left( S^\dagger \frac{d S}{dE}\right)=\frac{1}{\pi} \frac{d\delta}{dE} \, .
\label{spectraltheory21}		
\end{eqnarray}
If the $S$-matrix is diagonal, then in each one-dimensional subspace we obtain
\begin{eqnarray}
	S(E)= e^{\iimg \delta(E)} \, , \qquad \rho(E) = \frac{1}{\pi} \frac{d\delta(E)}{dE} \, .
\label{spectraltheory20}		
\end{eqnarray}

\subsection{Determinants and Ultralocality \label{Ultra}}
 The quadratic fluctuations    of a  single boson $\{ x(\t) \}$ give a Gaussian path integral
  \be \label{path}
 Z = \int [\mathcal{D}x(\t)] \, e^{-  \frac{1}{2\b}  \langle x | A x \rangle }\, .
 \ee
 where the inner product over the field space and the operator $A$ are defined by
 \be
\langle x | y \rangle :=  \int_0^\beta\, d\tau x(\t)   Y (\t) \, , \qquad  A =  -\frac{d^{2}}{d\t^{2}}
 \ee
 The  field $x(\t)$ can be expanded in terms of the modes of the operator 
 $A$   on the base circle:
 \be
 x(\t) = \sum_{n\in \bZ} e^{\frac{2\pi i n\tau}{\beta}} x_{n} \, , \qquad x^{*}_{n} = x_{-n} \, .
 \ee
The Gaussian integral can then be written as
 \be \label{path1}
Z= \int dx_0 \prod_{n>0}\left[\int dx_{-n}\,  dx_{n}
 \exp\left[ -\beta\left(\frac{2\pi n}{\beta} \right)^2 x_{-n}x_n  \right]
 \right] =
 \int dx_0  \prod_{n >  0}\left[\frac{{\beta}^{2}}{4\pi n^2} 
 \right]
 \ee
The infinite product can be regularized using  a zeta function
\begin{eqnarray}
\zeta(s) = \sum_{n = 1}^{\infty} \frac{1}{n^s} = \frac{1}{\Gamma(s)} \int_0^\infty \frac{x^{s-1}}{e^x - 1} \, dx
\label{zetadefn1}
\end{eqnarray}
Another formula that would be useful is the following 
\begin{eqnarray}
\frac{d}{ds} \zeta(s)=\frac{d}{ds}\sum_{n=1}^\infty e^{-s\, \ln\,  n}	= -\sum_{n=1}^\infty e^{-s\, \ln\,  n}\ln\,  n
\label{zetadefn2}
\end{eqnarray}
Hence we get,
 \bea
 \ln\,  Z &=&	\sum_n \ln \left(\frac{2\pi n^2}{\beta}\right)
 =\ln \left(\frac{2\pi }{\beta}\right)\sum_n 1-2\sum_n \ln \left(n\right)
\eea
Using  \eqref{zetadefn1} the first summation gives $\zeta(0)$ and using  \eqref{zetadefn2} the second summation gives $\zeta^\prime(0)$ 
\bea
 \ln \left(\frac{2\pi  }{\beta}\right)\zeta(0) -2 \zeta^\prime(0) = \frac{1}{2}\ln (2\pi \beta ) 
\eea
to obtain
 \begin{eqnarray}\label{Zdet}
  Z =	  \int \frac{dx_0}{\sqrt{2\pi \beta } } \frac{1}{\sqrt{\textrm{det}^\prime (A)}} 
 \end{eqnarray}
 where $\textrm{det}^\prime(A) $ is the renormalized determinant  of $A$ over nonzero modes. 
The factor of $\frac{1}{\sqrt{2 \pi \beta}}$ for a bosonic zero mode can be deduced more directly by using `ultralocality' as we explain below. Separating the path integral \eqref{path1} into integral over the zero mode $x_{0}$ and non-zero modes $x'$ of the operator $A$, we get
\be
\label{ultralocal}
\int \, dx_0 [dx'] e^{-\frac{1}{2\b }\langle x | A x \rangle} =  \int \, dx_0 [dx']  e^{-\frac{1}{2 \beta} \langle x' | A x' \rangle} = 
\frac{1}{\sqrt{\textrm{det}^\prime (A )}} \int \, dx_0 [dx']  e^{-\frac{1}{2 \beta} \langle x | x \rangle}
\ee
We now  to set the normalization by setting
\be
\int \, [d x] e^{-\frac{1}{2 \beta} \langle x |  x \rangle}  =1 \, .
\ee
It follows from the fact that the  measure for this integral involves only the local metric  on the base (and not its derivatives). Hence, it can be normalized to unity by an ultralocal counter-term corresponding to the `cosmological constant'. We then obtain
\be
\int \,  [dx']  e^{-\frac{1}{2 \beta} \langle x | x \rangle} = \frac{1}{\int dx_0 \exp ({-\frac{x^{2}_0 }{2 \beta} })}
 = \frac{1}{\sqrt{2\pi \b}} \, .
\ee
Substituting  in \eqref{ultralocal} we reproduce \eqref{Zdet}. This argument makes it apparent that for each bosonic zero mode we get a factor of $\frac{1}{\sqrt{2\pi \b}}$.

\bibliographystyle{JHEP}
\bibliography{eta}
\end{document}